\documentclass[floats,floatfix,prd,twocolumn,superscriptaddress,nofootinbib,longbibliography]{revtex4-2}

\pdfoutput=1
\usepackage{amsfonts,amssymb,amsmath}
\usepackage[usenames,dvipsnames]{xcolor}
\definecolor{292}{rgb}{0.3843, 0.6588, 0.8980}
\usepackage{scrextend}
\usepackage{tablefootnote}
\usepackage[unicode, colorlinks=true,
linkcolor=blue,
urlcolor=blue,
citecolor=292]{hyperref}
\usepackage[all]{hypcap}
\usepackage{graphicx}
\usepackage{xspace}
\usepackage{tikz}
\usepackage{float}
\usepackage{enumerate}
\usetikzlibrary{shapes,arrows}
\usetikzlibrary{matrix}
\usetikzlibrary{shapes.multipart}
\usetikzlibrary{er,positioning}
\usepackage[normalem]{ulem} 
\usepackage{url}
\usepackage{color,soul}
\usepackage[caption=false]{subfig}
\usepackage[T1]{fontenc}
\usepackage{lmodern}
\usepackage[utf8]{inputenc}
\usepackage{microtype}
\usepackage{array}
\newcolumntype{C}[1]{>{\centering\let\newline\\\arraybackslash\hspace{0pt}}m{#1}}
\usepackage{makecell}
\usepackage{diagbox}

\graphicspath{{plots/}}

\renewcommand{\d}[2]{\frac{d #1}{d #2}}

\usepackage{orcidlink}

\newcommand{\Cornell}{\affiliation{Cornell Center for Astrophysics and Planetary Science,
		Cornell University, Ithaca, New York 14853, USA}}
\newcommand{\Caltech}{\affiliation{Theoretical Astrophysics 350-17,
		California Institute of Technology, Pasadena, California 91125, USA}}
\newcommand{\OleMiss}{\affiliation{Department of Physics and Astronomy,
		University of Mississippi, University, Mississippi 38677, USA}}
\newcommand{\PennState}{\affiliation{Institute for Gravitation and the Cosmos \& Physics Department, Penn State, University Park, Pennsylvania 16802, USA}}
\newcommand{\MaxPlanck}{\affiliation{Max Planck Institute for Gravitational Physics (Albert Einstein Institute), Am M{\"u}hlenberg 1, Potsdam 14476, Germany}}

\begin{document}

\preprint{APS/123-QED}

\title{Fixing the BMS Frame of Numerical Relativity Waveforms}
\author{{Keefe Mitman}
\orcidlink{0000-0003-0276-3856}}
\email{kmitman@caltech.edu}
\Caltech
\author{Neev Khera
\orcidlink{0000-0003-3515-2859}}
\PennState
\author{Dante A. B. Iozzo
\orcidlink{0000-0002-7244-1900}}
\Cornell
\author{Leo C. Stein
\orcidlink{0000-0001-7559-9597}}
\OleMiss
\author{{Michael Boyle}
\orcidlink{0000-0002-5075-5116}}
\Cornell
\author{Nils Deppe
\orcidlink{0000-0003-4557-4115}}
\Caltech
\author{Lawrence E. Kidder
\orcidlink{0000-0001-5392-7342}}
\Cornell
\author{Jordan Moxon
\orcidlink{0000-0001-9891-8677}}
\Caltech
\author{\\Harald P. Pfeiffer
\orcidlink{0000-0001-9288-519X}}
\MaxPlanck
\author{Mark A. Scheel
\orcidlink{0000-0001-6656-9134}}
\Caltech
\author{Saul A. Teukolsky
\orcidlink{0000-0001-9765-4526}}
\Caltech
\Cornell
\author{William Throwe
\orcidlink{0000-0001-5059-4378}}
\Cornell

\hypersetup{pdfauthor={Mitman et al.}}

\date{\today}

\begin{abstract}
  \noindent Understanding the Bondi-Metzner-Sachs (BMS) frame of the
  gravitational waves produced by numerical relativity is crucial for
  ensuring that analyses on such waveforms are performed properly.
  It is also important that models are built from waveforms in the same BMS
  frame. Up until now, however, the BMS frame of numerical
  waveforms has not been thoroughly examined, largely because the necessary tools have not existed. In this paper, we show how to analyze and map to
  a suitable BMS frame for numerical waveforms calculated with the
  Spectral Einstein Code (SpEC). However, the methods and tools that we present
  are general and can be applied to any numerical waveforms. We
  present an extensive study of 13 binary black hole systems that broadly span parameter space. From these
  simulations, we extract the strain and also the Weyl
  scalars using both SpECTRE's Cauchy-characteristic
  extraction module and also the standard
  extrapolation procedure with
  a displacement memory correction applied during postprocessing. First,
  we show that the current center-of-mass correction used to map
  these waveforms to the center-of-mass frame is not as effective
  as previously thought. Consequently, we also develop an
  improved correction that utilizes asymptotic Poincar\'e charges
  instead of a Newtonian center-of-mass trajectory. Next, we map our
  waveforms to the post-Newtonian (PN) BMS frame using a PN strain
  waveform. This
  helps us find the unique BMS transformation that
  minimizes the $L^{2}$ norm of the difference
  between the numerical
  and PN strain waveforms during the early inspiral phase. We find that
  once the
  waveforms are mapped to the PN BMS frame, they can be
  hybridized with a PN strain waveform much more effectively than
  if one used any of the previous alignment schemes,
  which only utilize the Poincar\'e transformations.
\end{abstract}

\maketitle

\tableofcontents

\section{Introduction}
\label{sec:introduction}
As more and more astrophysical events are observed by gravitational wave detectors such as LIGO,\footnote{The Laser Interferometer Gravitational-Wave Observatory.} Virgo, and KAGRA,\footnote{The Kamioka Gravitational Wave Detector.} accurate models of gravitational waves for these systems are indispensable for conducting precise parameter estimation as well as tests of general relativity. As of now, the most accurate models of gravitational waves are the waveforms from numerical relativity (NR) simulations, with the most prevalent being those that correspond to binary black hole (BBH) mergers~\cite{SXSCatalog,Boyle:2019kee,Jani:2016wkt,Healy:2017psd}.

Like any observable system in nature, however, the waveforms produced by numerical relativity simulations are functions of the frame that they are in. Currently, the Simulating eXtreme Spacetimes (SXS) Catalog~\cite{SXSCatalog}---the largest of the publicly available waveform catalogs---and the RIT Catalog~\cite{Healy:2020vre} attempt to fix the Poincar\'e frame of their asymptotic waveforms by mapping them to the unique center-of-mass (CoM) frame through what they call a \emph{center-of-mass correction}~\cite{SXSCatalog,Boyle:2019kee,Woodford:2019tlo,Healy:2020vre}.\footnote{This fixing of the Poincar\'e frame was called a ``correction'' in the work of Ref.~\cite{Woodford:2019tlo} because the BBH system's center-of-mass drift is an unexpected phenomenon in numerical relativity simulations. It appears to be related
to an imperfect boundary condition on the gauge degrees of freedom at
the outer boundary.}

This correction---or this fixing of the Poincar\'e frame---uses the masses and the trajectories of the two black holes to construct the Newtonian center-of-mass trajectory
\begin{align}
\label{eq:newtcom}
\vec{x}_{\text{CoM}}\equiv\frac{m_{a}}{M}\vec{x}_{a}+\frac{m_{b}}{M}\vec{x}_{b},
\end{align}
where $m_{a}$ and $m_{b}$ are the Christodoulou masses~\cite{Christodoulou:1970wf} of the primary and secondary black holes, $M\equiv m_{a}+m_{b}$ is the total initial mass of the BBH system, and $\vec{x}_{a}$ and $\vec{x}_{b}$ describe the motion of the black holes' centers, i.e., the simulation coordinate averages over each one of the black hole's apparent horizon. With this trajectory, the translation and boost that best map the system to the Newtonian center-of-mass frame can then be found by finding the transformation that minimizes the average square of the distance between the center-of-mass and the origin of the corrected frame~\cite{Woodford:2019tlo}.

While this method for fixing the Poincar\'e frame has proven useful, there are certain aspects of this correction that are concerning. For one, when more post-Newtonian (PN) terms are included in the Newtonian calculation of the center of mass in Eq.~\eqref{eq:newtcom}, the correction does not improve and, for higher mass-ratio or precessing systems, becomes noticeably worse~\cite{Woodford:2019tlo}. Apart from this, it is also worrisome that the frame of these waveforms is being fixed based on information from the bulk of spacetime, rather than information from the waveforms themselves, especially as it has been shown that the coordinate velocity of the black holes does not accurately correspond to the asymptotic velocity for complicated systems~\cite{Iozzo2021}.

A better way to fix the Poincar\'e frame is to measure the Poincar\'e charges for asymptotic quantities and try to find the Poincar\'e transformation that changes these charges in a prescribed way. For example, to map a general relativistic system to the center-of-mass frame, one could compute the linear momentum charge and the boost charge from which the system's boost and translation away from the origin can be established. Or, as we will do in this work, one could instead compute just the center-of-mass charge and then determine which Poincar\'e transformation minimizes this charge.

Apart from the Poincar\'e frame, however, there is an additional freedom in general relativity arising from the extra symmetries of asymptotically flat spacetimes that extends the usual Poincar\'e group. This infinite group of symmetries, which was found by Bondi, van der Burg, Metzner, and Sachs~\cite{Bondi:1962px,Sachs:1962wk}, is known as the \emph{BMS group}.\footnote{There is also a proposed generalization of the BMS group, which promotes the Lorentz transformations to be the infinite group of local diffeomorphisms on $S^{2}$~\cite{Campiglia:2014yka,Campiglia:2015yka}. In this work, however, we will focus on just the BMS group and reserve an examination of the generalized BMS group for future study.} Fundamentally, the BMS group is a semidirect product of the usual Lorentz group with an infinite-dimensional Abelian group of transformations called \emph{supertranslations}, which contain the more familiar spacetime translations as a normal subgroup.

The most straightforward way to understand how a supertranslation affects coordinates is via the following. First define the Bondi time $u\equiv t-r$. Under an arbitrary spacetime translation $(\delta t,\delta\vec{x})=(\delta t,\delta x,\delta y,\delta z)$, we can write the corresponding transformation of $u$ as
\begin{align}
\label{eq:spacetimetranslation}
u'=u-\sum\limits_{\ell<2}\sum\limits_{m\leq|\ell|}\alpha_{\ell m}Y_{\ell m}(\theta,\phi),
\end{align}
where
\begin{subequations}
	\begin{align}
	\alpha_{0,0}&=\sqrt{4\pi}\delta t,\\
	\alpha_{1,\pm1}&=-\sqrt{\frac{2\pi}{3}}(\mp\delta x+i\delta y),\\
	\quad\alpha_{1,0}&=-\sqrt{\frac{4\pi}{3}}\delta z.
	\end{align}
\end{subequations}
A proper supertranslation $\alpha(\theta,\phi)$, i.e., a supertranslation that is not one of the spacetime translations (see Eq.~\eqref{eq:spacetimetranslation}), then acts on $u$ as
\begin{align}
u'=u-\alpha(\theta,\phi)
\end{align}
for
\begin{align}
\label{eq:supertranslation}
\alpha=\sum\limits_{\ell\geq2}\sum\limits_{m\leq|\ell|}\alpha_{\ell m}Y_{\ell m}(\theta,\phi)
\end{align}
with $\alpha_{\ell,m}=(-1)^{m}\overline{\alpha}_{\ell,-m}$ to make sure that $u'$ is real. Consequently, a supertranslation can be understood as a direction-dependent time translation on the boundary of asymptotically flat spacetimes, e.g., future null infinity. For example, if there exists a network of observers on a sphere surrounding a source, then ideally they could combine their received signals with some understanding of their clocks' synchronization. At future null infinity, such a synchronization becomes impossible and we could supply a separate time offset, i.e., a supertranslation, to each observer without changing the observable physics. An outline of how supertranslations transform the typical gravitational wave quantities, such as the strain $h$, the news $\dot{h}$, and the Weyl scalars $\Psi_{i}$, can be found in~\cite{Boyle:2015nqa,GomezLopez:2017kcw}.

There are really only two reasonable ways to fix this supertranslation freedom. The first, and simplest, is to find the supertranslation that minimizes the difference between a NR waveform and a PN waveform. The second, which is often the most common in the literature~\cite{Moreschi:1988pc,Moreschi:1998mw, Dain:2002mj}, can be understood as follows.

Like fixing the Poincar\'e frame by making use of the Poincar\'e charges, a fairly similar scheme can be executed to fix a BBH system's proper supertranslation freedom by utilizing the proper supertranslation charge. We refer to the frame that fixes this supertranslation freedom as the Bondi frame, while the frame that corresponds to fixing all of the freedom of our waveforms is called the BMS frame. Put differently, the BMS frame captures both the Poincaré and the proper supertranslation freedom, whereas the Bondi frame only involves the proper supertranslations. Just as the Bondi four-momentum is the charge that is related to spacetime translations, the \emph{supermomentum}---an infinite extension of the usual Bondi four-momentum---is the charge that corresponds to supertranslations. Thus, the supertranslation freedom can also be uniquely fixed by finding the supertranslation that minimizes the proper supermomentum charge. The Bondi frame under this transformation is then related to the Poincar\'e frame that corresponds to the transformation that minimizes the three-momentum, i.e., the rest frame. In this case, though, we call the Bondi frame with minimal\footnote{The reason we say minimal rather than no supermomentum is because of a subtlety having to do with gravitational memory, which we discuss in the Appendix~\ref{sec:superrestframe}.} supermomentum the \emph{nice section}~\cite{Dain:2002mj} or the \emph{super rest frame}.

In this work, we do exactly this. That is, we fix the Poincar\'e frame of our asymptotic waveforms by working with particular Poincar\'e charges to map the BBH systems to the center-of-mass frame. We then also fix the proper supertranslation freedom by finding the supertranslation that maps the waveforms to the Bondi frame that is ideal for current observations. Thus, we not only improve upon the work of~\cite{Woodford:2019tlo} by using relativistic charges instead of Newtonian trajectories, but we also wholly fix the proper supertranslation freedom, thereby fixing the complete BMS frame of our numerical waveforms.

Most importantly, we find that the new and improved fixing of the Poincar\'e frame is a drastic improvement over the previous method that uses Newtonian trajectories. Based on our observations, we conclusively find that the previous center-of-mass correction appears to have only approximately worked for the equal mass nonspinning, aligned spins, and superkick systems, and fails for the nonequal mass, anti-aligned spins, or precessing systems. Further, even in the equal mass nonspinning systems where the previous correction was nearly the same as the charge-based one, this new method nonetheless shows obvious improvements, such as reducing the leakage of the strain $(2,2)$ mode into other, subdominant strain modes. We show an example of this improvement in Fig.~\ref{fig:ModeChange}.

Apart from improving the fixing of the Poincar\'e frame, we also make a few important observations regarding the fixing of the supertranslation freedom. Even though it is often mentioned that the supertranslation freedom can be fixed by mapping to the super rest frame~\cite{Moreschi:1988pc,Moreschi:1998mw,Dain:2002mj}, we find that the most practical way to fix the supertranslations is by mapping our BBH systems to the PN Bondi frame. This is because LIGO expects numerical waveforms to agree with PN waveforms during the early inspiral phase of a BBH merger. Note, however, that for conducting any kind of analyses on quasinormal modes, the preferred BMS frame is actually the super rest frame, since this is the frame expected by the Teukolsky formalism.\footnote{This will be covered in a future work.}

Previously, the BMS freedom of numerical waveforms has not been important because the SXS Collaboration's waveforms did not exhibit the displacement memory~\cite{Mitman:2020pbt}. However, because waveforms with memory effects can now be easily produced by numerical relativity~\cite{Mitman:2020pbt} or can even have memory effects added to them via a correction~\cite{Mitman:2020bjf}, fixing this Bondi frame has become crucial, seeing as it is absolutely necessary for performing hybridizations between numerical and PN strain waveforms. We find that by mapping numerical waveforms to the PN BMS frame, we can significantly improve NR/PN strain hybridizations. Our main result regarding this is shown in Fig.~\ref{fig:PNAlignmentErrors}.

Last, we also discover that by completely fixing the BMS frame of our waveforms, we can perform noticeably better convergence tests of numerical relativity waveforms, since waveforms from different resolutions can now be compared while in the same BMS frame.

\subsection{Overview}
\label{sec:overview}

We organize our computations and results as follows.\\In Sec.~\ref{sec:ellleq2} we introduce the four main Poincar\'e charges that are useful when examining asymptotic waveforms: the linear momentum, angular momentum, boost, and center-of-mass charges. Apart from this, we also discuss how the center-of-mass charge can be used to obtain the system's velocity and translation away from the origin. Following this, in Sec.~\ref{sec:ellgeq2} we discuss the two most natural Bondi frames and conclude that for practical purposes, the most useful frame to map to is the PN Bondi frame. Finally, in Secs.~\ref{sec:ellleq2results} and~\ref{sec:ellgeq2results} we present our results for mapping to the center-of-mass frame as well as the PN BMS frame and thus illustrate how the BMS frame of numerical waveforms should be fixed for future analyses and, most importantly, future surrogate models~\cite{Blackman:2017dfb,Blackman:2017pcm,Varma:2018mmi,Williams:2019vub}.

\subsection{Conventions}
\label{sec:conventions}

We set $c=G=1$ and take $\eta_{\mu\nu}$ to be the $(-,+,+,+)$ Minkowski metric. When working with complex dyads, following the work of~\cite{Iozzo:2020jcu}, we use
\begin{align}
\label{eq:dyads}
q_{A}=-\frac{1}{\sqrt{2}}(1,i\sin\theta)\text{ and }q^{A}=-\frac{1}{\sqrt{2}}(1,i\csc\theta),
\end{align}
and write the round metric on the two-sphere $S^{2}$ as $q_{AB}$. The complex dyad obeys the following properties
\begin{align}
q_{A}q^{A}=0,\quad q_{A}\bar{q}^{A}=1,\quad q_{AB}=q_{A}\bar{q}_{B}+\bar{q}_{A}q_{B}.
\end{align}
Note that this convention differs from the related works of~\cite{Mitman:2020bjf,Mitman:2020pbt,Moxon:2020gha}, which in contrast do not include the $1/\sqrt{2}$ normalization factor on the dyads in Eq.~\eqref{eq:dyads}. We choose this convention because it makes our expressions for the asymptotic charges in Eq.~\eqref{eq:charges} more uniform. Nonetheless, for transparency we provide the conversion between our quantities and those of these previous works in Eq.~\eqref{eq:conversion}. We build spin-weighted fields with the dyads as follows. For a tensor field $W_{A\cdots D}$, the function
\begin{align}
W=W_{A\cdots BC\cdots D}q^{A}\cdots q^{B}\bar{q}^{C}\cdots\bar{q}^{D}
\end{align}
with $m$ factors of $q$ and $n$ factors of $\bar{q}$ has a spin weight of $s=m-n$. When raising and lowering spin weights we use the Geroch-Held-Penrose differential spin-weight operators $\eth$ and $\bar{\eth}$~\cite{GHP1973},
\begin{subequations}
	\begin{align}
	\eth W&=(D_{E}W_{A\cdots BC\cdots D})q^{A}\cdots q^{B}\bar{q}^{C}\cdots\bar{q}^{D}q^{E},\\
	\bar{\eth}W&=(D_{E}W_{A\cdots BC\cdots D})q^{A}\cdots q^{B}\bar{q}^{C}\dots\bar{q}^{D}\bar{q}^{E}.
	\end{align}
\end{subequations}
Here, $D_{A}$ is the covariant derivative on the two-sphere. The $\eth$ and $\bar{\eth}$ operators in spherical coordinates are then
\begin{subequations}
	\begin{align}
	\eth W(\theta,\phi)&=-\frac{1}{\sqrt{2}}(\sin\theta)^{+s}(\partial_{\theta}+i\csc\theta\partial_{\phi})\nonumber\\
	&\phantom{=.-\frac{1}{\sqrt{2}}(\sin(\theta))^{s}}\left[(\sin\theta)^{-s}W(\theta,\phi)\right],\\
	\bar{\eth} W(\theta,\phi)&=-\frac{1}{\sqrt{2}}(\sin\theta)^{-s}(\partial_{\theta}-i\csc\theta\partial_{\phi})\nonumber\\
	&\phantom{=.-\frac{1}{\sqrt{2}}(\sin(\theta))^{s}}\left[(\sin\theta)^{+s}W(\theta,\phi)\right].
	\end{align}
\end{subequations}
Thus, when acting on spin-weighted spherical harmonics, these operators produce
\begin{subequations}
	\begin{align}
	\eth(\phantom{}_{s}Y_{\ell m})&=+\frac{1}{\sqrt{2}}\sqrt{(\ell-s)(\ell+s+1)}_{s+1}Y_{\ell m},\\
	\bar{\eth}(\phantom{}_{s}Y_{\ell m})&=-\frac{1}{\sqrt{2}}\sqrt{(\ell+s)(\ell-s+1)}_{s-1}Y_{\ell m}.
	\end{align}
\end{subequations}
We denote the gravitational wave strain\footnote{We explicitly define the strain as described in Appendix C of~\cite{Boyle:2019kee}.} by $h$, which we represent in a spin-weight $-2$ spherical harmonic basis,
\begin{align}
h(u,\theta,\phi)=\sum\limits_{\ell,m}h_{\ell m}(u)\,{}_{-2}Y_{\ell m}(\theta,\phi),
\end{align}
where, again, $u\equiv t-r$ is the Bondi time. We denote the Weyl scalars by $\Psi_{0-4}$. The conversion from the convention of~\cite{Moxon:2020gha,Mitman:2020pbt,Mitman:2020bjf} (denoted $\texttt{NR}$\footnote{$\texttt{NR}$ because this is the convention that corresponds to the outputs of the SXS simulations.}) to ours (denoted $\texttt{MB}$\footnote{$\texttt{MB}$ because this corresponds to the Moreschi-Boyle convention used in the works~\cite{Iozzo:2020jcu,Boyle:2015nqa,Moreschi1986} and the code \texttt{scri}~\cite{scri_url,Boyle:2013nka,Boyle:2014ioa,Boyle:2015nqa}.}) is
\begin{align}
\label{eq:conversion}
h^{\texttt{NR}}=2\bar{\sigma}^{\texttt{MB}}\quad\text{and}\quad\Psi_{i}^{\texttt{NR}}=\frac{1}{2}(-\sqrt{2})^{i}\Psi_{i}^{\texttt{MB}}.
\end{align}
Note that we will omit these superscripts and henceforth assume that everything is in the $\texttt{MB}$ convention.

\section{Fixing the $\ell<2$ Transformations}
\label{sec:ellleq2}

As discussed in the Introduction, for the past few years the method for fixing the Poincar\'e frame of BBH systems in numerical relativity has relied on using the Newtonian center-of-mass trajectory, i.e., Eq.~\eqref{eq:newtcom}~\cite{Woodford:2019tlo,Boyle:2019kee,Healy:2020vre}. While this has served as a successful first step, we can improve upon this by using certain Poincar\'e charges: specifically, the center-of-mass charge.

We first present the main asymptotic Poincar\'e charges. These charges are the linear momentum charge $P_{\Psi}$, the angular momentum charge $J_{\Psi}$, the boost charge $K_{\Psi}$, and the energy moment charge $E_{\Psi}$. Others have just called $E_{\Psi}$ the center-of-mass charge, but this is misleading because $E_{\Psi}$ really measures the center of mass scaled by the energy of the system. Thus, we instead refer to $E_{\Psi}$ as the energy moment charge and reserve $G_{\Psi}$ to represent the center-of-mass charge. These charges are computed by integrating the Bondi mass aspect $m$, the Lorentz aspect $N$, and the energy moment aspect $E$, which are derived in the $\texttt{NR}$ convention in Appendixes A and B of~\cite{Mitman:2020pbt}. In the $\texttt{MB}$ convention these are
\begin{subequations}
\begin{align}
m&\equiv-\text{Re}\left[\Psi_{2}+\sigma\dot{\bar{\sigma}}\right],\\
N&\equiv-\left(\Psi_{1}+\sigma\eth\bar{\sigma}+u\eth m+\frac{1}{2}\eth\left(\sigma\bar{\sigma}\right)\right),\\
E&\equiv N+u\eth m=-\left(\Psi_{1}+\sigma\eth\bar{\sigma}+\frac{1}{2}\eth\left(\sigma\bar{\sigma}\right)\right).
\end{align}
\end{subequations}
Consequently, the main Poincar\'e charges are
\begin{subequations}
\label{eq:charges}
\begin{align}
P_{\Psi}&=\frac{1}{4\pi}\int_{S^{2}}\Psi\,m\,d\Omega,\\
J_{\Psi}&\equiv\frac{1}{4\pi}\int_{S^{2}}\epsilon^{AB}\left(\partial_{A}\Psi\right)\left(\bar{q}_{B}N\right)\,d\Omega,\nonumber\\
&=\frac{1}{4\pi}\int_{S^{2}}\text{Re}\left[\left(\bar{\eth}\Psi\right)\left(-iN\right)\right]\,d\Omega,\\
K_{\Psi}&\equiv\frac{1}{4\pi}\int_{S^{2}}q^{AB}\left(\partial_{A}\Psi\right)\left(\bar{q}_{B}N\right)\,d\Omega,\nonumber\\
&=\frac{1}{4\pi}\int_{S^{2}}\text{Re}\left[\left(\bar{\eth}\Psi\right)N\right]\,d\Omega,\\
\label{eq:energymomentcharge}
E_{\Psi}&\equiv\frac{1}{4\pi}\int_{S^{2}}q^{AB}\left(\partial_{A}\Psi\right)\left(\bar{q}_{B}\left(N+u\eth m\right)\right)\,d\Omega,\nonumber\\
&=\frac{1}{4\pi}\int_{S^{2}}\text{Re}\left[\left(\bar{\eth}\Psi\right)\left(N+u\eth m\right)\right]\,d\Omega,\nonumber\\
&=K_{\Psi}+uP_{\Psi}.
\end{align}
\end{subequations}
where $\epsilon^{AB}$ is the usual Levi-Civita tensor and $\Psi$ is a real spin-weight $0$ function on the two-sphere. This scalar $\Psi$ is typically taken to be a unique combination of the $\ell\leq1$ spherical harmonics so as to represent one of the four Cartesian coordinates $t,x,y,z$, i.e.,
\begin{subequations}
	\begin{align}
	t&=1\nonumber\\
	&=\sqrt{4\pi}Y_{(0,0)},\\
	x&=\sin\theta\cos\phi\nonumber\\
	&=\sqrt{\frac{4\pi}{3}}\left[\frac{1}{\sqrt{2}}\left(Y_{(1,-1)}-Y_{(1,+1)}\right)\right],\\
	y&=\sin\theta\sin\phi\nonumber\\
	&=\sqrt{\frac{4\pi}{3}}\left[\frac{i}{\sqrt{2}}\left(Y_{(1,-1)}+Y_{(1,+1)}\right)\right],\\
	z&=\cos\theta\nonumber\\
	&=\sqrt{\frac{4\pi}{3}}Y_{(1,0)}.
	\end{align}
\end{subequations}
By utilizing some properties of the spherical harmonics, we can create a four-vector from the projection of a charge along each Cartesian direction:
\begin{subequations}
	\label{eq:chargeCart}
	\begin{align}
	\label{eq:timecomponent}
	A^{t}&=\frac{1}{\sqrt{4\pi}}A_{(0,0)},\\
	A^{x}&=\frac{1}{\sqrt{4\pi}}\frac{1}{\sqrt{6}}\text{Re}\left[A_{(1,-1)}-A_{(1,+1)}\right],\\
	A^{y}&=\frac{1}{\sqrt{4\pi}}\frac{1}{\sqrt{6}}\text{Im}\left[A_{(1,-1)}+A_{(1,+1)}\right],\\
	A^{z}&=\frac{1}{\sqrt{4\pi}}\frac{1}{\sqrt{3}}\text{Re}\left[A_{(1,0)}\right],
	\end{align}
\end{subequations}
where $A_{(\ell,m)}$ is the $(\ell,m)$ mode of the aspect $A$ when $A$ is written in a spin-weight $0$ or $1$ spherical harmonic basis.\footnote{Note that $A=m$ is a spin-weight $0$ function and $A=-iN$, $N$, and $N+u\eth m$ are spin-weight $1$ functions.} Note that all of these charges and aspects, which we have defined in Eq.~\eqref{eq:charges}, have been previously examined in earlier works such as~\cite{Dray:1984rfa,Dray:1984gz,Flanagan:2015pxa,Compere:2019gft,Mitman:2020pbt}.

As mentioned in the introduction of this section, we are primarily interested in the center-of-mass charge $G_{\Psi}$, which is closely related to the energy moment charge $E_{\Psi}$ in Eq.~\eqref{eq:energymomentcharge}. By definition, the center-of-mass charge is the energy moment divided by the energy of the system,
\begin{align}
\label{eq:centerofmasscharge}
G_{\Psi}&\equiv\frac{E_{\Psi}}{P^{t}}=\frac{E_{\Psi}}{\gamma M_{B}},
\end{align}
where $P^{t}$ is computed according to Eq.~\eqref{eq:timecomponent} with $A=m$, $\gamma$ is the Lorentz factor
\begin{align}
\gamma=\sqrt{1-\left|\vec{P}/P^{t}\right|^{2}}^{-1},
\end{align}
and $M_{B}$ is the Bondi mass~\cite{Bondi:1962px}
\begin{align}
M_{B}\equiv\sqrt{-\eta_{\mu\nu}P^{\mu}P^{\nu}}.
\end{align}
The charge $G_{\Psi}$ measures the center of mass that evolves linearly as a function of momentum in the absence of gravitational radiation~\cite{Compere:2019gft}. The reason why we are mainly interested in this Poincar\'e charge is because
\begin{subequations}
\label{eq:Gproperties}
\begin{align}
\vec{G}|_{u=0}&=\left(\vec{K}/\gamma M_{B}\right)|_{u=0},\\
\dot{\vec{G}}&=\d{}{u}\left[\left(\vec{K}+u\vec{P}\right)/\left(\gamma M_{B}\right)\right]\nonumber\\
&=\vec{v}+u\dot{\vec{v}}-\frac{\vec{K}}{\gamma M_{B}}\left[\frac{\dot{M}_{B}}{M_{B}}-\gamma^{2}\left(\vec{v}\cdot\dot{\vec{v}}\right)\right]+\frac{\dot{\vec{K}}}{\gamma M_{B}}\nonumber\\
&\approx\vec{v}+\text{(oscillations about $\vec{v}$)},
\end{align}
\end{subequations}
since
\begin{align}
\label{eq:assumptions}
\dot{M}_{B}\approx0,\quad\dot{\vec{v}}\approx0,\quad\text{and}\quad\dot{\vec{K}}\approx0,
\end{align}
in the inspiral and ringdown phases of the BBH merger.\footnote{We address issues with this assumption in Sec.~\ref{sec:ellleq2results}.} Therefore, the intercept of $\vec{G}\equiv(E^{x},E^{y},E^{z})/P^{t}$ will be the center of mass of the system at $u=0$ and the slope will be the three-velocity. This means that by fitting a degree one polynomial to this three-vector, we can obtain the amount a system is translated and boosted out of the center-of-mass frame and then apply the opposite Poincar\'e transformation to map our waveforms to the center-of-mass frame.\footnote{While we could also use the boost and linear momentum charges to obtain these transformations, computing both of these charges is more computationally expensive than if we just compute the center-of-mass charge, since there are fewer products of waveforms that need to be taken when finding $G_{\Psi}$.} Note that these Poincar\'e charges in Eq.~\eqref{eq:charges} can also be used to measure properties of the BBH system or the remnant black hole, as was illustrated in the recent and related work of \cite{Iozzo2021}.

\section{Fixing the $\ell\geq2$ Transformations}
\label{sec:ellgeq2}

We now move on to a discussion about the proper supertranslation freedom.

\subsection{Mapping to the super rest frame}
\label{sec:ellgeq2superrestframe}

While it is often mentioned that a system's Bondi frame should be fixed by minimizing the supermomentum~\cite{Moreschi:1998mw},\\ current observatories expect their waveform models to resemble PN expansions. Therefore, even though fixing a system's Bondi frame using the supermomentum is a well-motivated option with unique benefits, we reserve a discussion of this for Appendix~\ref{sec:superrestframe}, since none of the results that we present involve this technique.

\begin{table*}
	\label{tab:runs}
	\centering
	\renewcommand{\arraystretch}{1.2}
	\begin{tabular}{@{}l@{\hspace*{7mm}}c@{\hspace*{7mm}}c@{\hspace*{7mm}}c@{}c@{}c@{}c@{\hspace*{7mm}}c@{}c@{}c@{}c@{}}
		\Xhline{3\arrayrulewidth}
		Name & CCE radius & $q$ & $\chi_{A}$:\, & $(\hat{x},\,$ & $\hat{y},\,$ & $\hat{z})$ & $\chi_{B}$:\, & $(\hat{x},\,$ & $\hat{y},\,$ & $\hat{z})$\\
		\hline
		\texttt{q1\_nospin}              & 292 & $1.0$ & & $(0,\,$ & $0,\,$ &  $0)$ & & $(0,\,$ & $0,\,$ & $0)$ \\
		\texttt{q1\_aligned\_chi0\_2}    & 261 & $1.0$ & & $(0,\,$ & $0,\,$ & $0.2)$ & & $(0,\,$ & $0,\,$ & $0.2)$ \\
		\texttt{q1\_aligned\_chi0\_4}    & 250 & $1.0$ & & $(0,\,$ & $0,\,$ & $0.4)$ & & $(0,\,$ & $0,\,$ & $0.4)$ \\
		\texttt{q1\_aligned\_chi0\_6}   & 236 & $1.0$ & & $(0,\,$ & $0,\,$ & $0.6)$ & & $(0,\,$ & $0,\,$ & $0.6)$ \\
		\texttt{q1\_antialigned\_chi0\_2} & 274 & $1.0$ & & $(0,\,$ & $0,\,$ & $0.2)$ & & $(0,\,$ & $0,\,$ & $-0.2)$ \\
		\texttt{q1\_antialigned\_chi0\_4} & 273 & $1.0$ & & $(0,\,$ & $0,\,$ & $0.4)$ & & $(0,\,$ & $0,\,$ & $-0.4)$ \\
		\texttt{q1\_antialigned\_chi0\_6} & 270 & $1.0$ & & $(0,\,$ & $0,\,$ & $0.6)$ & & $(0,\,$ & $0,\,$ & $-0.6)$ \\
		\texttt{q1\_precessing}          & 305 & $1.0$ & & $(0.487,\,$ & $0.125,\,$ & $-0.327)$ & & $(-0.190,\,$ & $0.051,\,$ & $-0.227)$ \\
		\texttt{q1\_superkick}           & 270 & $1.0$ & & $(0.6,\,$ & $0,\,$ & $0)$ & & $(-0.6,\,$ & $0,\,$ & $0)$ \\
		\texttt{q4\_nospin}              & 235 & $4.0$ & & $(0,\,$ & $0,\,$ & $0)$ & & $(0,\,$ & $0,\,$ & $0)$ \\
		\texttt{q4\_aligned\_chi0\_4}    & 222 & $4.0$ & & $(0,\,$ & $0,\,$ & $0.4)$ & & $(0,\,$ & $0,\,$ & $0.4)$ \\
		\texttt{q4\_antialigned\_chi0\_4} & 223& $4.0$ & & $(0,\,$ & $0,\,$ & $0.4)$ & & $(0,\,$ & $0,\,$ & $-0.4)$ \\
		\texttt{q4\_precessing}           & 237 & $4.0$ & & $(0.487,\,$ & $0.125,\,$ & $-0.327)$ & & $(-0.190,\,$ & $0.051,\,$ & $-0.227)$ \\
		\Xhline{3\arrayrulewidth}
	\end{tabular}
	\caption{Parameters of the BBH mergers used in our results. The mass ratio is $q=M_A/M_B$, and the initial dimensionless spins of the two black holes are $\chi_A$ and $\chi_B$. These simulations have been made publicly available at~\cite{ExtCCECatalog}.}
\end{table*}

\subsection{Mapping to the PN BMS frame}
\label{sec:ellgeq2PNBondiFrame}

Besides using the supermomentum to map a system to its super rest frame, the supertranslation freedom of waveforms can also be fixed by mapping them to their corresponding PN Bondi frame. Unlike PN waveforms, the NR waveforms are finite in length and do not contain information from the BBH system's entire past history. As a result, when numerical strain waveforms that contain memory are created, either by using Cauchy-characteristic extraction (CCE)~\cite{Pollney:2010hs,Mitman:2020pbt} or by correcting extrapolated waveforms~\cite{Mitman:2020bjf}, their average during the inspiral phase will tend to be close to zero. By contrast, post-Newtonian waveforms typically have a memory contribution that is monotonically increasing with time and only approaches zero as $u\rightarrow-\infty$. Therefore, if hybridizations of the numerical waveforms are to be made with PN waveforms, a mapping to the PN Bondi frame is essential to ensure that the waveforms and their memory contributions can be properly aligned. While we examined the results of mapping our systems to the super rest frame, we find that, because of this hybridization concern, mapping to the PN Bondi frame is the more sensible procedure for fixing the Bondi frame. Consequently, in Sec.~\ref{sec:results} we only present the results for mapping our various BBH systems to their corresponding PN BMS frame, as defined by a PN strain waveform, and reserve a study of the benefits of mapping to the super rest frame for future work.

\section{Results}
\label{sec:results}

For the following results, we numerically evolved a set of 13 binary black hole mergers with various mass ratios and spin configurations using the Spectral Einstein Code (SpEC)~\cite{SpECCode}. We list the important parameters of these various BBH systems in Table~\ref{tab:runs}. Each simulation contains roughly 19 orbits prior to merger and is evolved until the waves from ringdown leave the computational domain. Unlike the evolutions in the SXS catalog, the full set of Weyl scalars and the strain have been extracted from these runs and the waveforms have been computed using the extrapolation technique described in~\cite{Iozzo:2020jcu} and the CCE procedure described in~\cite{Moxon:2020gha,moxon2021}. Extrapolation is performed with the python module \texttt{scri}~\cite{scri_url, Boyle:2013nka, Boyle:2015nqa, Boyle:2014ioa} and CCE is run with SpECTRE's CCE module~\cite{Moxon:2020gha,moxon2021,CodeSpECTRE}.

For the CCE extractions, the four world tubes that are available have radii that are equally spaced between $2\lambdabar_{0}$ and $21\lambdabar_{0}$, where $\lambdabar_0=1/\omega_0$ is the initial reduced gravitational wavelength as determined by the orbital frequency of the binary from the initial data. Based on the recent work of~\cite{Mitman:2020bjf}, however, we choose to use only the waveforms that correspond to the world tube with the second-smallest radius, since these waveforms have been shown to minimally violate the BMS balance laws. For clarity, we provide the world tube radius used for each system in Table~\ref{tab:runs}. All of these 13 BBH systems' waveforms have been made publicly available at~\cite{ExtCCECatalog}.

As mentioned above, the asymptotic strain waveforms are computed using two methods: extrapolation and CCE. The first method utilizes Regge-Wheeler-Zerilli (RWZ) extraction to compute the strain on a series of concentric spheres of constant coordinate radius and then proceeds to extrapolate these values to future null infinity $\mathcal{I}^{+}$ using $1/r$ approximations~\cite{Sarbach:2001qq,Regge:1957td,Zerilli:1970se,Boyle:2019kee,Iozzo:2020jcu, Boyle:2009vi}.  This is the strain that can be found in the SXS catalog. The other and more faithful extraction method, which is known as CCE, computes the strain by using the world tube data provided by a Cauchy evolution as the inner boundary data for a nonlinear evolution of the Einstein field equations on null hypersurfaces~\cite{Moxon:2020gha,moxon2021}. CCE requires freely specifying the strain on the initial null hypersurface labeled $u=0$. Like~\cite{Mitman:2020pbt,Mitman:2020bjf}, we choose this field to match the value and the first radial derivative of $h$ from the Cauchy data on the world tube using the ansatz
\begin{align}
h(u=0,r,\theta^{A})=\frac{A(\theta^{A})}{r}+\frac{B(\theta^{A})}{r^{3}},
\end{align}
where the two coefficients $A(\theta^{A})$ and $B(\theta^{A})$ are fixed by the Cauchy data on the world tube. Unfortunately, constructing a satisfactory initial null hypersurface for CCE is currently an open issue in numerical relativity. Consequences of this choice manifest as transient effects arising at early times~\cite{Mitman:2020pbt}. We address these in Sec.~\ref{sec:transienteffects}.

As for the extrapolated strain waveforms that we use, these have been postprocessed so that they exhibit the displacement memory effect and are thus more on par with the waveforms produced by CCE~\cite{Mitman:2020bjf}.

Last, when performing our analysis, we predominantly use the code \texttt{scri}~\cite{scri_url,Boyle:2013nka,Boyle:2014ioa,Boyle:2015nqa} to compute BMS charges and transform asymptotic waveform quantities.

\subsection{$\ell<2$ Results}
\label{sec:ellleq2results}

\begin{figure}
	\label{fig:ChargeExample}
	\centering
	\includegraphics[width=\columnwidth]{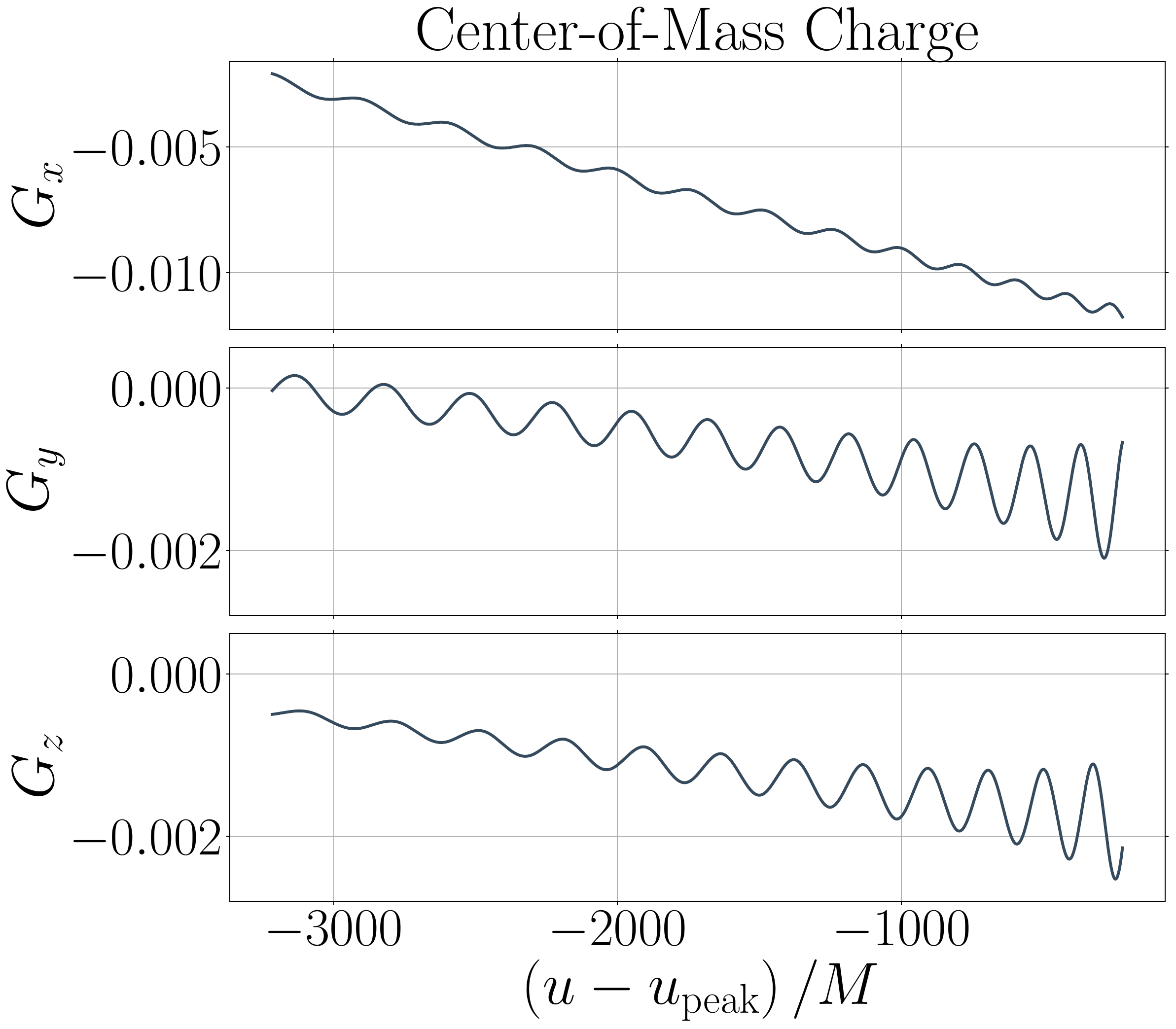}
	\caption{Example of what the center-of-mass charge looks like for a
		system with large center-of-mass velocity;
		in this case a $q=4$, precessing system. The charges plotted are computed directly from Eq.~\eqref{eq:centerofmasscharge}, where $\vec{E}$ and $P^{t}$ are obtained from Eq.~\eqref{eq:chargeCart} by using the energy moment aspect $E\equiv N+u\eth m$ and mass aspect $m$. We define the peak time to be the peak of the $L^{2}$ norm of the Geroch supermomentum [see Eq.~\eqref{eq:Geroch}], since this quantity is a supertranslation-invariant quantity~\cite{Ashtekar_1981,Dain:2002mj}. The waveform used is a CCE waveform.\\
		BBH merger: \texttt{q4\_precessing} (see Table~\ref{tab:runs}).}
\end{figure}

As discussed in Sec.~\ref{sec:ellleq2}, a BBH system can be mapped to its center-of-mass frame by utilizing the center-of-mass charge $\vec{G}$. In Fig.~\ref{fig:ChargeExample}, we show this charge for a $q=4$, precessing BBH system (see Table~\ref{tab:runs}). We compute the plotted charges from Eq.~\eqref{eq:centerofmasscharge} by using a CCE waveform. As can be seen, in the $\hat{x}$, $\hat{y}$, and $\hat{z}$ directions the average of the center-of-mass charge is not constant with respect to the Bondi time. Further, if one imagines tracing these curves back in time then it can easily be observed that they begin with a nonzero value. Because of these results, we may assert that, with time, the BBH system is drifting through space away from a point that is not the origin. If the system under consideration were not drifting and were in the center-of-mass frame, then we would expect our charges to have both zero slope and zero intercept. Fortunately, because of the nature of this charge, to map to the center-of-mass frame one can simply boost and translate the system by the negative of the charge's slope and intercept. In the subsequent discussion, we first check to see if any of our of 13 binary systems, either before or after the Newtonian center-of-mass correction~\cite{Woodford:2019tlo}, are in the center-of-mass frame. After this, we then proceed to apply our charge-based center-of-mass correction and evaluate the improvement it has on our waveforms.

\subsubsection{A note on transient effects}
\label{sec:transienteffects}

Because numerical relativity simulations are evolved from imperfect initial data~\cite{Boyle:2019kee}, the output waveforms contain unphysical effects referred to as \emph{junk radiation}. In extrapolated waveforms, the junk radiation appears at early times and typically decays after an orbit or two. For CCE waveforms, however, the junk radiation tends to persist longer into the waveform~\cite{Mitman:2020pbt}. As a result, because we examine extrapolated and CCE waveforms, we perform our analysis on the part of our waveforms that is three orbits past the start of the simulation, since the CCE-specific transient effects have decayed by then. We refer to this time throughout the results as $u_{1}$.

\subsubsection{Determining the best method for fixing the Poincar\'e frame}
\label{sec:framefixingmethod}

\begin{figure}
	\label{fig:convergence}
	\centering
	\includegraphics[width=\columnwidth]{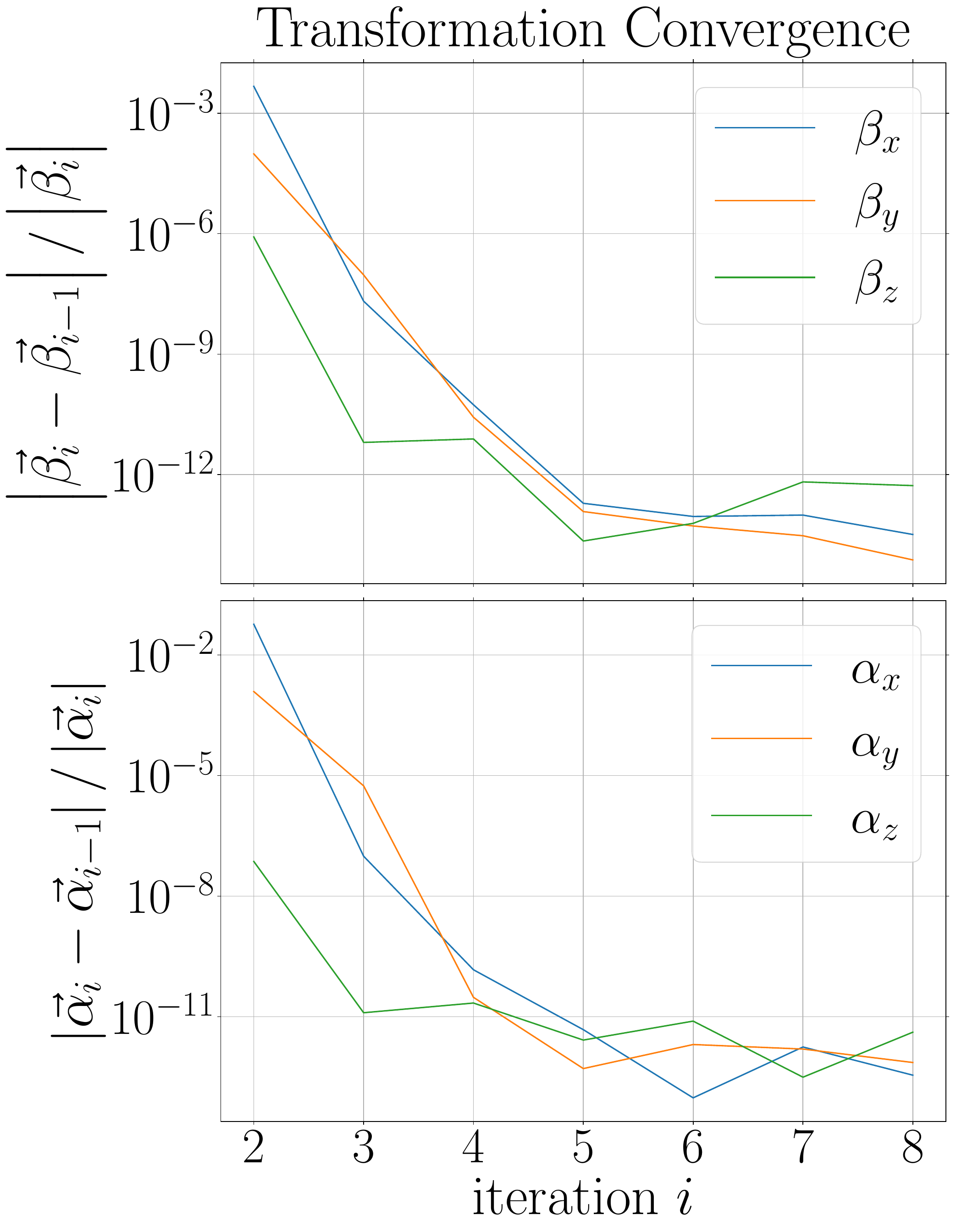}
	\caption{The convergence of the boost and translation vectors obtained by fitting to the center-of-mass charge. Here, $\beta_{i}$ ($\alpha_{i}$) represents the boost (translation) obtained after $i$ iterations of fitting to the charge $\vec{G}$, transforming the untransformed waveform with the fit result, fitting to the new charge, and then composing that fit with the previous vector. For example, $\beta_{1}$ is obtained from fitting to the charge once, and $\beta_{2}$ is obtained from fitting to the charge transformed using $\beta_{1}$ and then composing that fit with $\beta_{1}$. Note that we show the result for \texttt{q1\_aligned\_chi0\_6} because this simulation exhibits the slowest convergence out of the 13 systems that we examined. The waveform used is a CCE waveform.\\
		BBH merger: \texttt{q1\_aligned\_chi0\_6} (see Table~\ref{tab:runs}).}
\end{figure}

\begin{figure*}
	\label{fig:boostsandtranslations}
	\centering
	\includegraphics[width=\columnwidth]{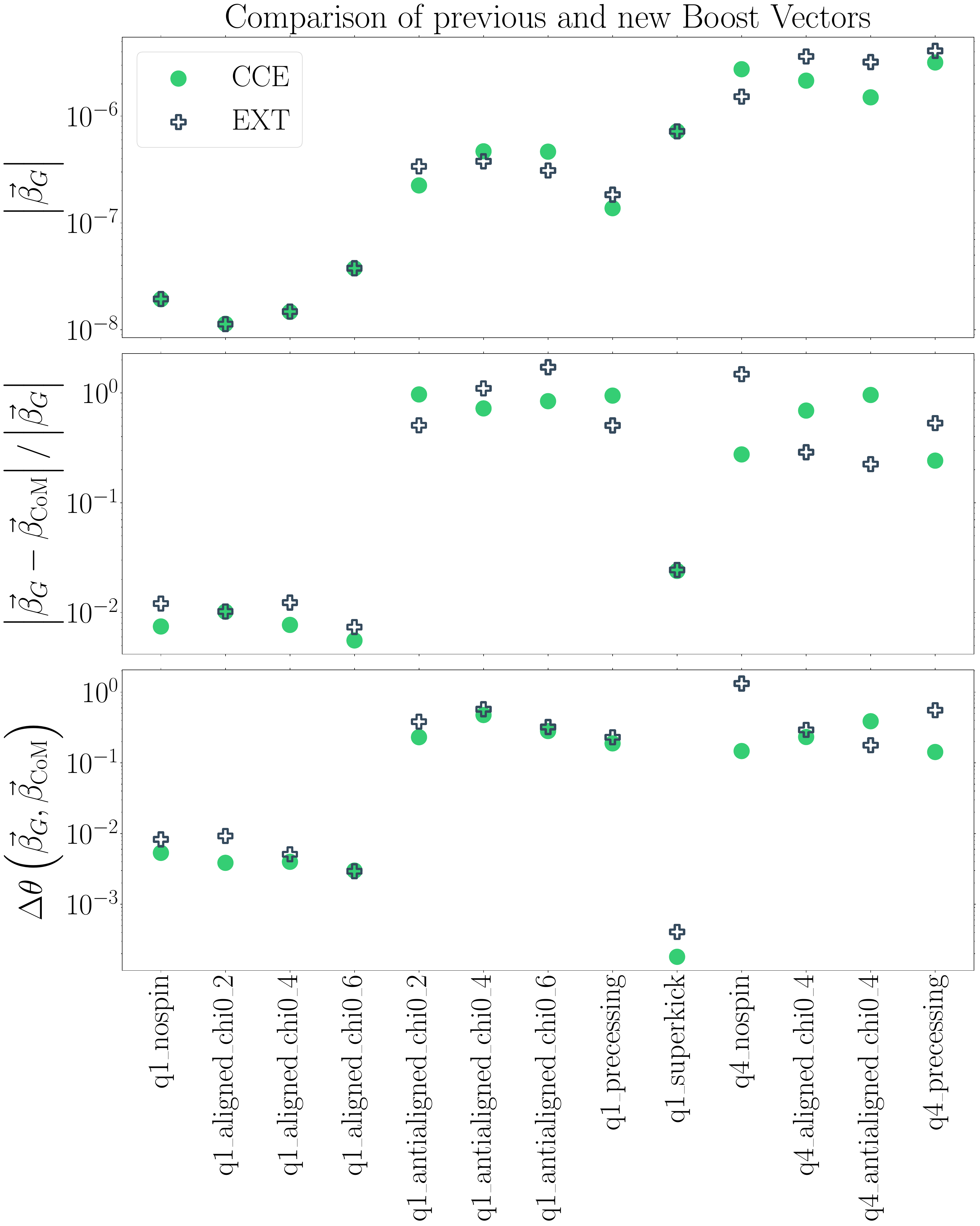}
	\includegraphics[width=\columnwidth]{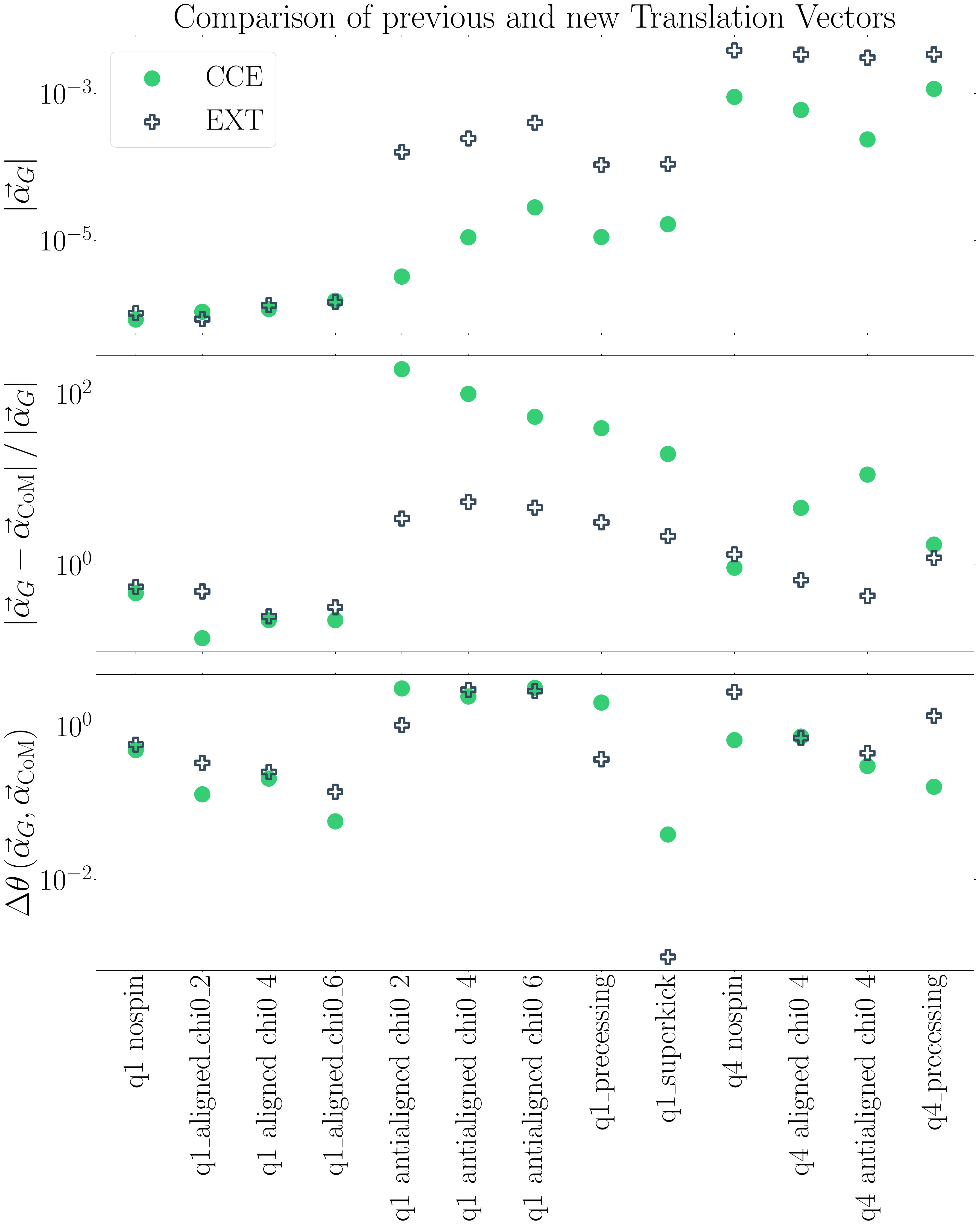}
  \caption{Examining the boost and translation vectors as measured by iteratively fitting to the center-of-mass charge $\vec{G}$, which is computed by using Eq.~\eqref{eq:chargeCart} to obtain the vector components of Eq.~\eqref{eq:centerofmasscharge}. In the top row we show the magnitude of these vectors, which we label $\vec{v}_{G}$ with $\vec{v}=\vec{\beta}$ for boosts (left) and $\vec{v}=\vec{\alpha}$ for translations (right). In the middle row we compute the relative error between these vectors and the vectors $\vec{v}_{\text{CoM}}$, which are obtained by the previous center-of-mass correction that relies on the Newtonian trajectories of the black holes. In the bottom row we plot the angle between these vectors using Eq.~\eqref{eq:angle}.}
\end{figure*}

We now compute the boosts and translations needed to map the 13 BBH systems to the center-of-mass frame. We first compute the charges according to Eq.~\eqref{eq:centerofmasscharge}. Next, we define an initial time $u_{1}$ to be three orbits past the start of the simulation and a final time that is three orbits before the peak time $u_{\text{peak}}$. We define the peak time to be the peak of the $L^{2}$ norm of the Geroch supermomentum [see Eq.~\eqref{eq:Geroch}], since this quantity is a supertranslation-invariant quantity~\cite{Ashtekar_1981,Dain:2002mj}. We choose this final time to ensure that we are only working with the inspiral phase of the binary, rather than the merger phase. Note that one could fix the Poincar\'e frame using the remnant BH, i.e., mapping the kick velocity to zero, but this is not as instinctive as using the inspiral phase to fix the frame, even though it would matter for fitting quasinormal modes. Equipped with the charges and boundary times, we then linearly fit to the center-of-mass charge in the $\hat{x}$, $\hat{y}$, and $\hat{z}$ directions and take the needed boost to be the negative of the slope and the needed translation to be the negative of the vertical intercept at $u=0$.

When we first applied this new Poincar\'e frame fixing, the improvements that we saw in the extreme systems, e.g., the $q=4$ and fast-spinning systems, were remarkable. For the simpler systems, however, the improvements were not as large as we expected them to be. This is because these systems are already reasonably close to being in the center-of-mass frame. Thus the measured boosts and translations are more susceptible to errors that are introduced by oscillations in the charge and also by the failure of the assumptions in Eq.~\eqref{eq:assumptions} to hold because of non-Newtonian effects. Fortunately, because this method for mapping to the center-of-mass frame just involves the computation of charges, it can be iterated. That is, after the initial transformation is found, it can be applied, and then the center-of-mass charge can be computed again from the new asymptotic waveform. With this new charge, a new transformation can then be found, which may be composed with the previous transformation to obtain a more accurate mapping to the center-of-mass frame.

In Fig.~\ref{fig:convergence} we plot the convergence of the boosts and translations for various iterations of this fitting procedure. We only show the results for \texttt{bbh\_q1\_aligned\_chi0\_6} because this system has the slowest convergence of the 13 binaries examined. In this figure, $\beta_{i}$ ($\alpha_{i}$) represents the boost (translation) obtained after $i$ iterations of this charge fitting process. More specifically, this procedure is as follows:
\begin{enumerate}[I.]
	\item Take $\vec{\beta}_{0},\vec{\alpha}_{0}=0$.
	\item If the iteration number $i$ is not $0$, transform the waveform with the Poincar\'e transformation $\vec{\beta}_{i},\vec{\alpha}_{i}$.
	\item Compute $\vec{G}$ from the transformed waveform.
	\item Obtain $\vec{\beta}$ and $\vec{\alpha}$ by fitting to $\vec{G}$ with a degree one polynomial ($\vec{\beta}=\text{slope}$, $\vec{\alpha}=\text{vertical intercept}$).
	\item Compute $\vec{\beta}_{i+1}=\vec{\beta}_{i}+\vec{\beta}$ and $\vec{\alpha}_{i+1}=\vec{\alpha}_{i}+\vec{\alpha}$.
	\item Repeat.
\end{enumerate}
As can be seen, even for this slowly converging system, the transformations converge rather quickly. Thus, based on this simulation, we choose a fixed number of five iterations for every one of our systems. With this many iterations, the improvements in the simpler systems then become as large as we would like to see.

\subsubsection{Comparing the Newtonian trajectory and charge-based frame fixing methods}
\label{sec:checkingthePoincareframe}

\begin{figure}
	\label{fig:ChargeChange}
	\centering
	\includegraphics[width=\columnwidth]{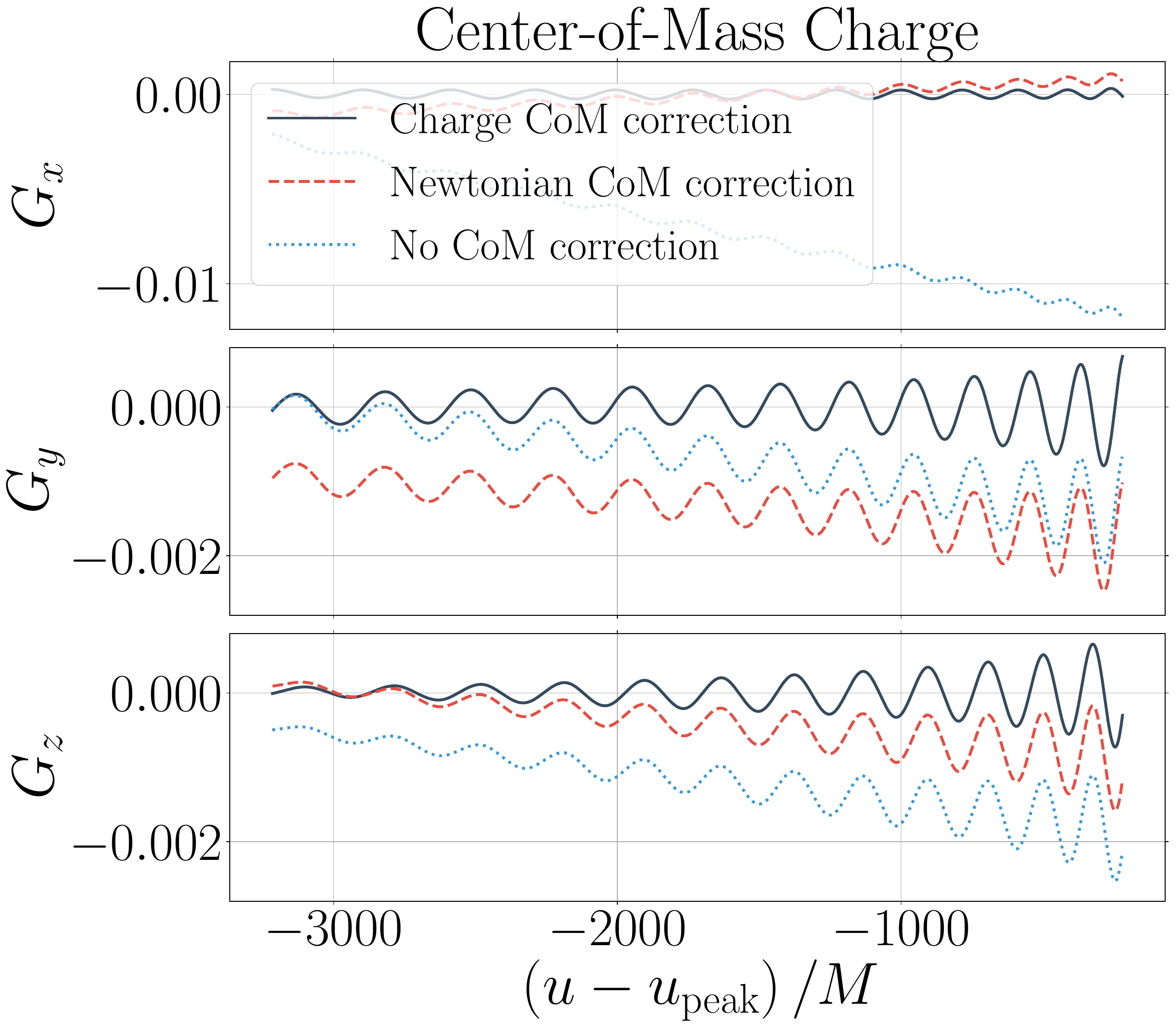}
	\caption{Comparing the center-of-mass charge from waveforms transformed using the Newtonian center-of-mass correction (red/dashed curves) to the charge obtained from those transformed using the charge-based method (black/solid curves). As a reference, we also plot the charge obtained from the untransformed waveforms (blue/dotted curves). The charges plotted are computed directly from Eq.~\eqref{eq:centerofmasscharge}, where $\vec{E}$ and $P^{t}$ are obtained from Eq.~\eqref{eq:chargeCart} by using the energy moment aspect $E\equiv N+u\eth m$ and mass aspect $m$. The waveform used is a CCE waveform.\\
		BBH merger: \texttt{q4\_precessing} (see Table~\ref{tab:runs}).\\}
\end{figure}

\begin{figure}
	\label{fig:ModeChange}
	\centering
	\includegraphics[width=\columnwidth]{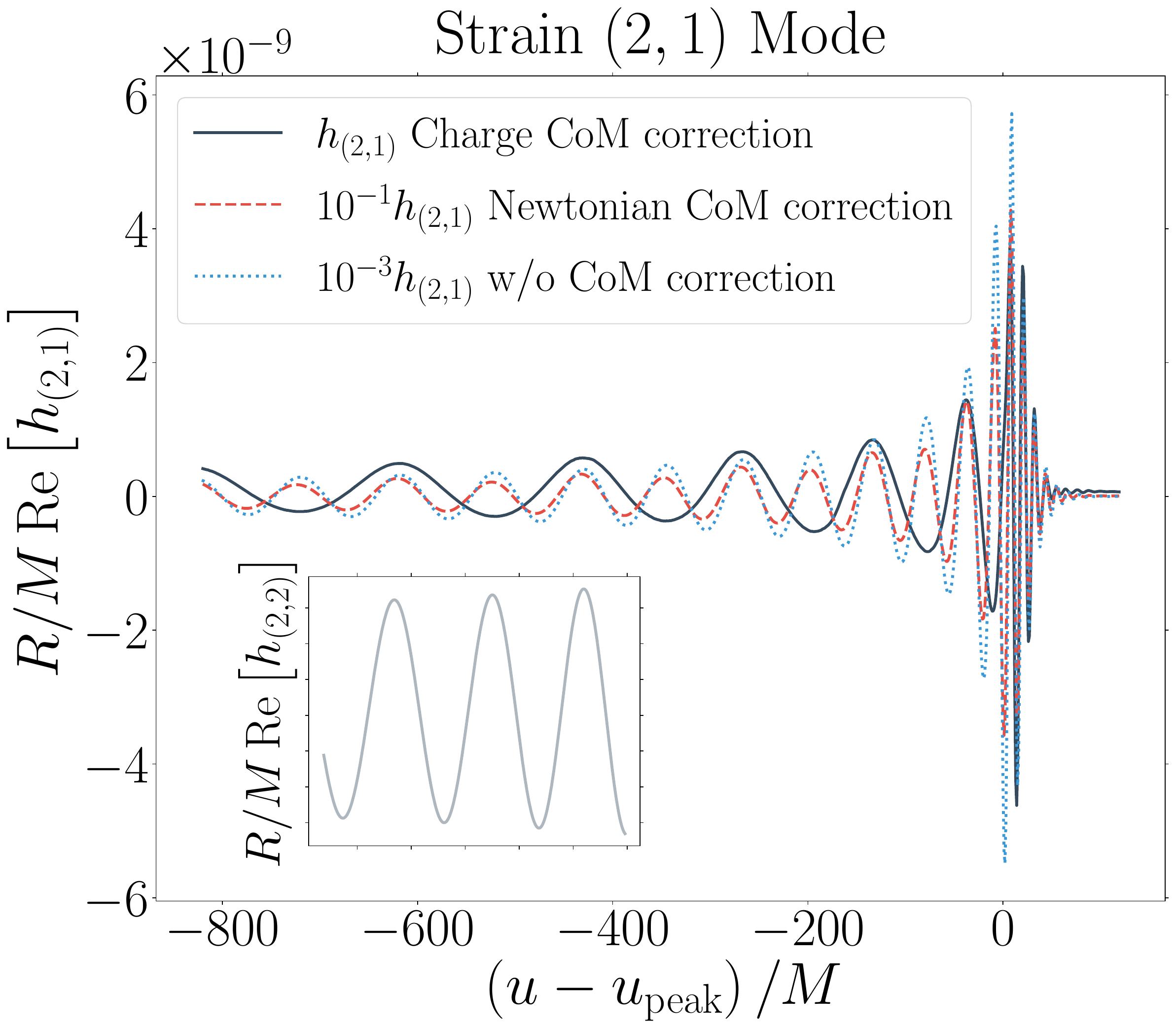}
  \caption{The strain $(2,1)$ mode of the original waveform (blue/dotted curve, scaled by $10^{-3}$) compared to that of the Newtonian (red/dashed curve, scaled by $10^{-1}$) and the charge-based (black/solid curve) center-of-mass corrections. In the inset plot we provide the strain $(2,2)$ mode of the original waveform to illustrate that unless the charge-based correction is used, the $(2,1)$ mode exhibits the same frequency as the $(2,2)$ mode. Note that the time axis of the inset plot matches up with the main plot's. The waveform used is a CCE waveform.\\
		BBH merger: \texttt{q1\_nospin} (see Table~\ref{tab:runs}).}
\end{figure}

\begin{figure}
	\label{fig:Mismatches}
	\centering
	\includegraphics[width=\columnwidth]{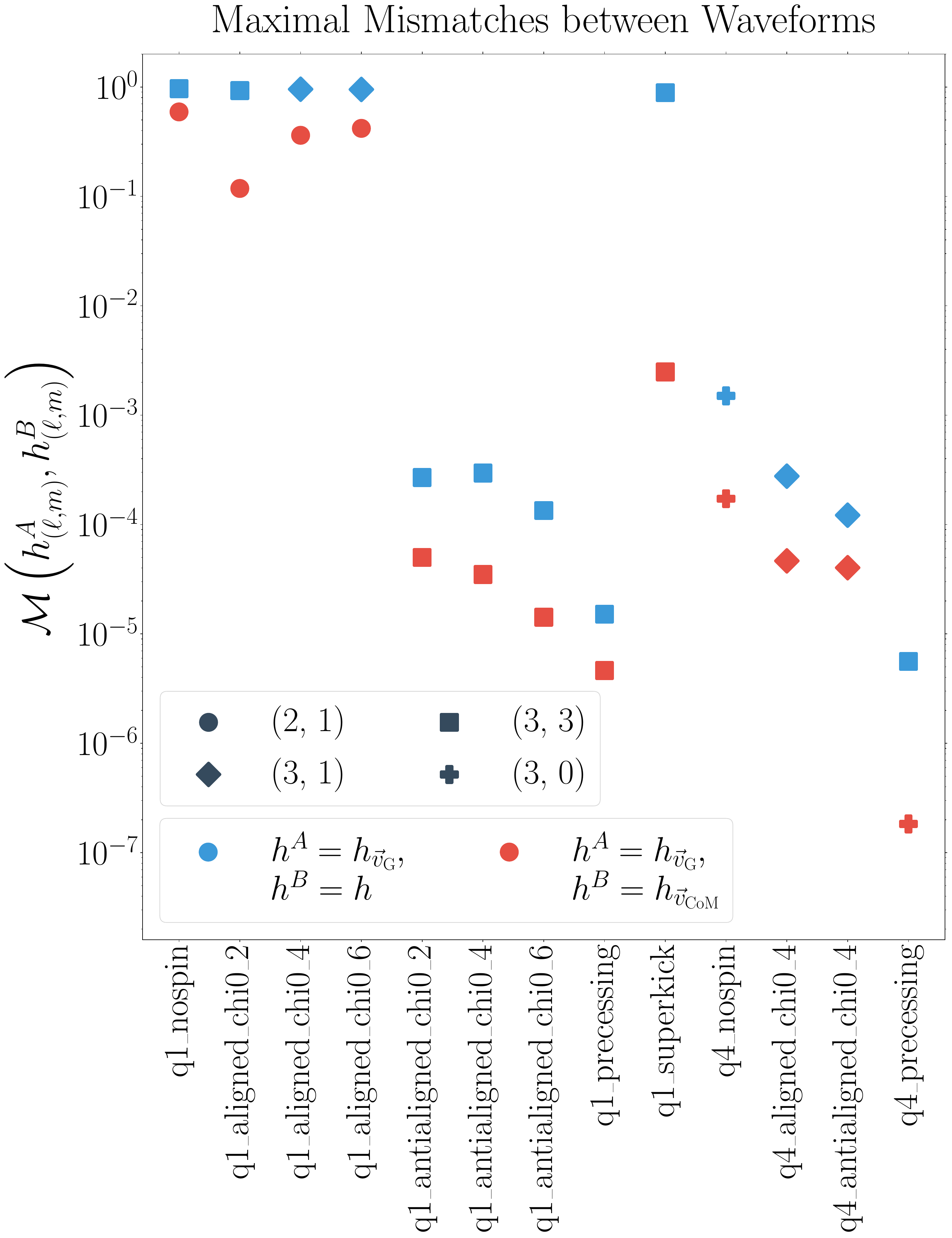}
	\caption{The modes that produce the largest mismatches between the charge-based center-of-mass corrected strains and both the original strains (blue) and the strains corrected using the Newtonian center-of-mass correction (red). The mismatch in each mode is computed by using Eq.~\eqref{eq:mismatch}. As shown in the plot's legend, the shape of each point represents the mode while the color represents the strains used in the mismatch. Note that the mismatches for the $q=1$ nonspinning and aligned-spin systems are so large because these systems have a rotation by $\pi$ symmetry, and thus when the charge-based center-of-mass correction is used the $m=\text{odd}$ modes become much closer to their expected value of zero. The waveforms used in these computations are CCE waveforms.}
\end{figure}

We now compare our iterative center-of-mass correction that uses the center-of-mass charge to the previous version that uses Newtonian trajectories. In Fig.~\ref{fig:boostsandtranslations} we show that of the 13 binaries examined, none of them are exactly in the center-of-mass frame after applying the correction that relies on Newtonian trajectories since all of the new Poincar\'e transformations are nonzero. More specifically, we plot two columns, one for the boosts on the left and one for the translations on the right. In the top row, we plot the magnitude of these vectors as obtained from fitting to the center-of-mass charge, which is a proxy for how much the system fails to be in the center-of-mass frame. In the middle row, we plot the relative difference between this charge-based vector $\vec{v}_{\text{G}}$, and the vector obtained from the Newtonian center-of-mass correction $\vec{v}_{\text{CoM}}$. This serves as a proxy for how much the system fails to be in the center-of-mass frame, even after the Newtonian correction. Finally, in the bottom row, we plot the angle between these two vectors. That is,
\begin{align}
\label{eq:angle}
\Delta\theta(\vec{v}_{\text{G}},\vec{v}_{\text{CoM}})\equiv\arccos\left(\frac{\vec{v}_{\text{G}}}{|\vec{v}_{\text{G}}|}\cdot\frac{\vec{v}_{\text{CoM}}}{|\vec{v}_{\text{CoM}}|}\right).
\end{align}
As can be seen in the top row, the equal mass nonspinning and aligned systems are reasonably close to being in the center-of-mass frame, while the other systems, especially the $q=4$ systems, are not. According to the other rows, though, because the differences between the vectors are so large, we realize that the Newtonian method for mapping to the center-of-mass frame is not nearly as successful as previously thought. Consequently, it is now evident that this Newtonian method for mapping a BBH system to its center-of-mass frame does not achieve its objective and the method based on the center-of-mass charge, which we explore in more detail now, is necessary.

\subsubsection{Examining improvements to waveforms}
\label{sec:waveformimprovements}

At this point, we examine how the strain waveforms change under the center-of-mass charge-based mapping to the center-of-mass frame. First, though, in Fig.~\ref{fig:ChargeChange} we show how the center-of-mass charge changes under the Newtonian center-of-mass correction versus the new charge-based method. What this plot shows is that while the Newtonian center-of-mass correction only corrects the $\hat{x}$ component of the center-of-mass charges for the \texttt{q4\_precessing} system, the new method produces an average value of exactly zero for every vector component.

In Fig.~\ref{fig:ModeChange} we show the most important consequence of improving the fixing of the Poincar\'e frame by using asymptotic Poincar\'e charges. For this example, we show the strain $(2,1)$ mode of \texttt{q1\_nospin} as it is, after the Newtonian center-of-mass transformation, and after our new center-of-mass transformation. We show this mode because it exhibits the largest mismatch
(see Eq.~\eqref{eq:mismatch}) when comparing strains that have been transformed using both the previous and the new center-of-mass corrections. Based on PN theory, during the system's inspiral phase we expect the frequency of this mode to be half the strain $(2,2)$ mode's frequency.\footnote{Really we expect this mode to be exactly zero because of the symmetry of the system; however, because the two spins and the eccentricity of the black holes in the \texttt{bbh\_q1\_nospin} simulation are not precisely zero, owing to numerical error, there is an unexpected nonzero contribution to this mode.} However, as can be seen by comparing the original and transformed waveforms, after correctly mapping to the center-of-mass frame the frequency of the strain $(2,1)$ mode is roughly half of what it was before. This is because, previously, the system was not truly in the center-of-mass frame so the strain $(2,2)$ mode was leaking into the $(2,1)$ mode. Note that in Fig.~\ref{fig:ModeChange} we have scaled the original waveform and the waveform transformed with the Newtonian center-of-mass correction by factors of $10^{-3}$ and $10^{-1}$ to make them more comparable to the waveform transformed with the charge-based center-of-mass correction.

Last, to show the impact this charge-based method has more broadly, we provide Fig.~\ref{fig:Mismatches}. This figure shows the mismatch between the newly transformed waveforms and both the original waveforms and the waveforms that have been transformed using the previous Newtonian correction. In this plot, we compute the mismatch between a mode of two strain waveforms via
\begin{align}
\label{eq:mismatch}
\mathcal{M}(&h_{(\ell,m)}^{A},h_{(\ell,m)}^{B})\equiv\nonumber\\
&\phantom{=.}1-\frac{\langle h_{(\ell,m)}^{A},h_{(\ell,m)}^{B}\rangle}{\sqrt{\langle h_{(\ell,m)}^{A},h_{(\ell,m)}^{A}\rangle\langle h_{(\ell,m)}^{B},h_{(\ell,m)}^{B}\rangle}},
\end{align}
in which the inner product is given by
\begin{align}
\langle h_{(\ell,m)}^{A},h_{(\ell,m)}^{B}\rangle\equiv\int_{u_{1}}^{+\infty}h_{(\ell,m)}^{A}\overline{h_{(\ell,m)}^{B}}du,
\end{align}
where $u=+\infty$ is the final time of the simulation.

\subsection{$\ell\geq2$ Results}
\label{sec:ellgeq2results}

\begin{figure*}
	\label{fig:PNAlignmentErrors}
	\centering
	\includegraphics[width=\textwidth]{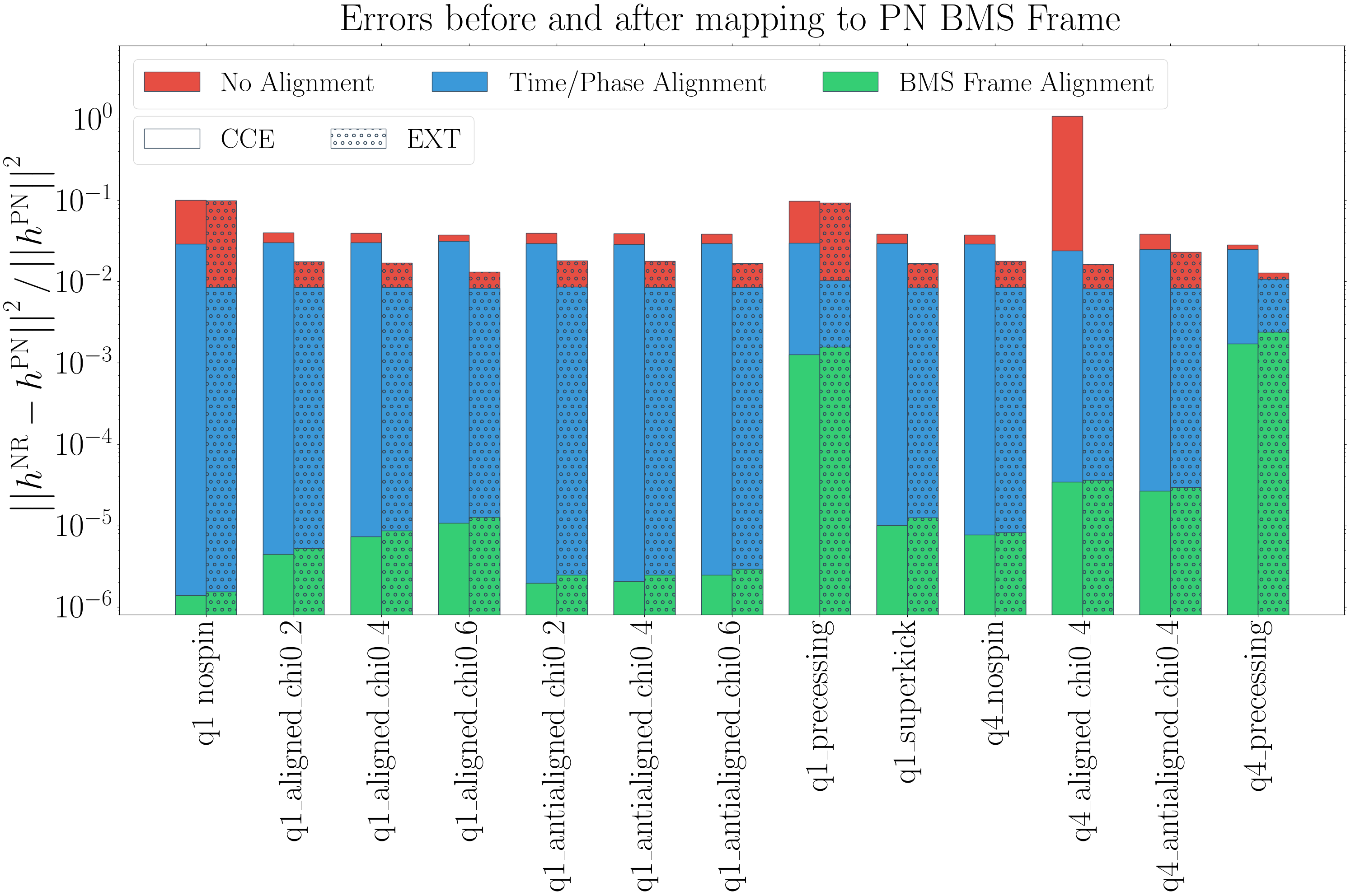}
	\caption{Comparing the BMS alignment results between NR and PN strain waveforms. In red, we show the initial misalignment. In blue, we show the misalignment if the usual alignment procedure is used, i.e., finding the time translation and frame rotation that best aligns a NR strain waveform that has undergone the previous center-of-mass correction to a PN strain waveform. Finally, in green, we show the misalignment after the new BMS frame alignment procedure has been used, that is, finding the BMS transformation (up to $\ell=4$) that minimizes the average $L^{2}$ norm of the difference of the NR and PN strain waveforms. Note that the measure of this misalignment, $||f(u,\theta,\phi)||^{2}$, is defined to be $||f(u,\theta,\phi)||^{2}=\int_{u_{1}}^{u_{1}+\text{four orbits}}\int_{S^{2}}f\bar{f}\,d\Omega\,du$.}
\end{figure*}

With the $\ell<2$ modes of our waveforms properly fixed using the new center-of-mass correction, we now explore how to fix the $\ell\geq2$ modes, i.e., choosing a Bondi frame. As discussed in Sec.~\ref{sec:ellgeq2}, there are really only two options: mapping the systems to their super rest frame or their PN Bondi frame. As described earlier, we prefer to map to the PN BMS frame since this tends to best improve the hybridization between two NR and PN strain waveforms.

To map our various systems to their PN BMS frame, we begin by creating a 3PN-order strain waveform from the orbital frequency of the two black holes using the code \texttt{GWFrames}~\cite{scri_url,Boyle:2013nka}. To generate this PN waveform, we obtain the orbital frequency of the system from the horizon information and evolve it backward in time using the PN evolution equations. We then simultaneously find the Poincar\'e transformation and the $2\leq\ell\leq4$ supertranslation that minimize the $L^{2}$ norm of the difference between the strain and the PN waveform. The norm is computed over the time interval starting at $u_1$ and continuing for four orbits. To perform this minimization, we use SciPy's minimize function corresponding to the sequential least squares programming algorithm (SLSQP)~\cite{Virtanen:2019joe} and define the following function:

\begin{enumerate}[I.]
	\item Take, as inputs, a NR strain, a PN strain, and also a center-of-mass transformation.
	\item Consider the ordered list of transformations
	\begin{enumerate}[1.]
		\item $\ell=2$ supertranslation,
		\item time translation,
		\item frame rotation,
		\item $\ell=3$ supertranslation,
		\item $\ell=4$ supertranslation.
	\end{enumerate}
	\item Begin with iteration $n=1$.
	\item For iteration $n$, include all transformations from the above list up to transformation $n$ as free parameters. Use the findings from iteration $n-1$ as initial guesses for the $n-1$ transformations.
	\item Use SciPy's SLSQP minimize function to find the collection of $n$ transformations that, when composed with the center-of-mass transformation (see below), best map the NR strain to the frame of the PN strain, i.e., the transformations that minimize the $L^{2}$ norm of the difference of the two waveforms, integrated over the time interval from $u_{1}$ to four orbits past $u_{1}$.
	\item Repeat until $n=5$.
\end{enumerate}

We find that it is important to start this procedure with the $\ell=2$ supertranslation, because this tends to be the largest source of error. Beyond this, however, the order of the transformations is fairly inconsequential and chosen as such to minimize run time. Note though that it is also important to allow the previous transformations to be free parameters in the next iteration because each new transformation tends to influence the previous ones.

With this function, we then run the iterative procedure:
\begin{enumerate}[I.]
	\item Find the center-of-mass transformation that maps the NR strain waveform to the center-of-mass frame using the charge-based center-of-mass correction.
	\item Provide this transformation and the NR and PN strain waveforms to the minimizing function.
	\item Apply the optimized BMS transformation to the raw NR strain\footnote{We apply the BMS transformation to the raw waveform to ensure that we are not transforming the same object more than once, which would introduce numerical error.} and find the center-of-mass transformation needed to map this new waveform to the center-of-mass frame.
	\item Compose this new transformation with the original center-of-mass transformation and then repeat steps II. - IV. with the previous BMS transformation as an initial guess until a desired precision is obtained.
\end{enumerate}

We find that by running this procedure 4 times we can obtain rather impressive alignments between the input NR and PN strain waveforms for most of our systems. We choose to run this method, rather than optimizing over all BMS transformations, because this method not only produces the best alignment but it also tends to keep the system much closer to its center-of-mass frame.

In Fig.~\ref{fig:PNAlignmentErrors}, we show the results of mapping the various CCE and extrapolated waveforms from our 13 systems to their corresponding PN BMS frame. In red, we show the initial misalignment. In blue, we show the misalignment if the usual alignment procedure is performed, i.e., finding the time translation and frame rotation that best aligns a PN strain waveform to a NR strain waveform that has undergone the previous center-of-mass correction. Finally, in green, we show the misalignment after using the new BMS frame alignment procedure. As is clearly illustrated, by capitalizing on the full BMS freedom of NR waveforms one can perform substantially better alignment between NR and PN strain waveforms. Apart from this, though, one can also observe the failure of the PN waveform to accurately model the BBH system, e.g., as the total spin or the mass ratio of the system increases, or if the system is precessing, the success of the BMS alignment between the NR and PN strains tends to worsen. This is expected, however, since the PN waveform is only of 3PN order. Regardless, this shows that not only can the Bondi frame, and really the whole BMS frame, be fixed by utilizing a PN strain waveform, but doing so is critically important for aligning, and thus hybridizing, NR and PN waveforms.

\begin{figure}
	\label{fig:hybrid}
	\centering
	\includegraphics[width=\columnwidth]{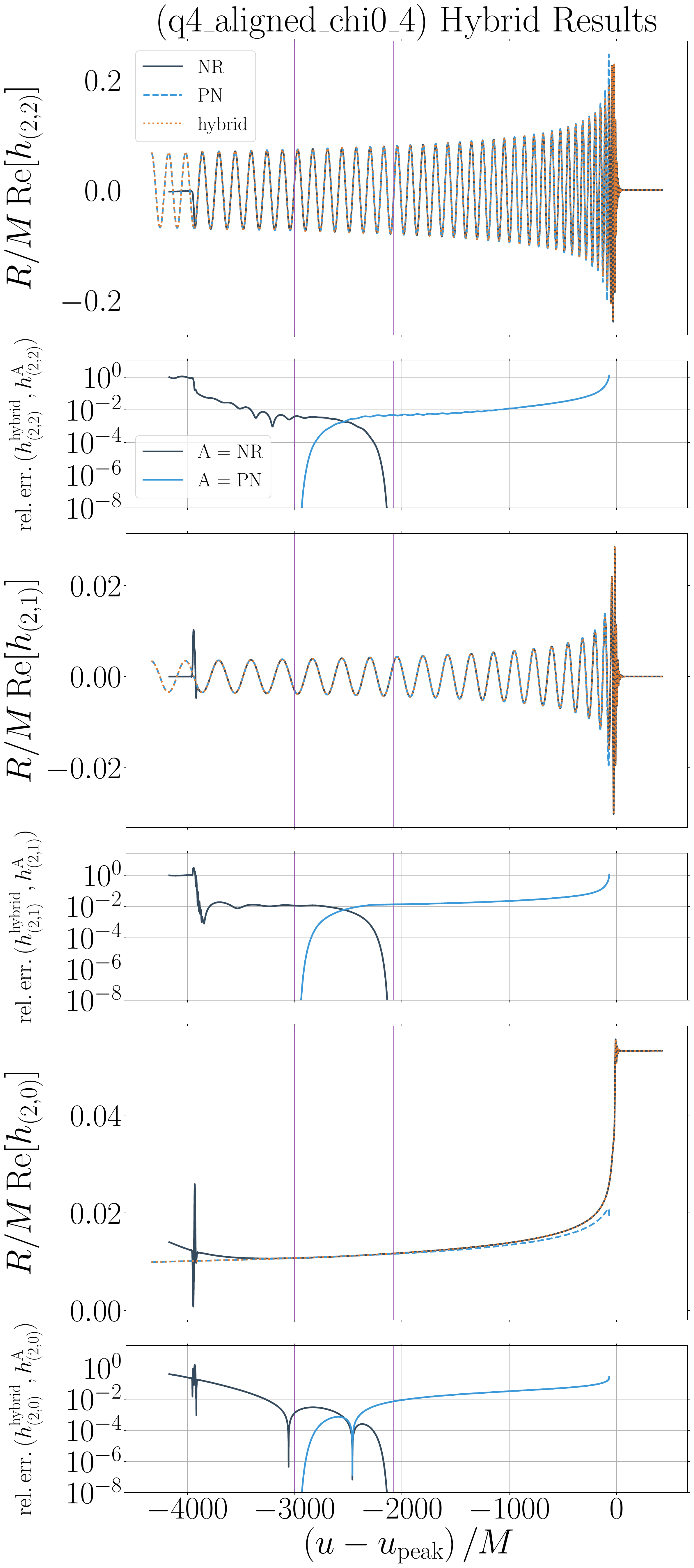}
	\caption{The NR and PN strain hybrid. We present three plots, one for each of the  $(2,2)$, $(2,1)$, and $(2,0)$ modes. In each plot, in the top panel we show the NR waveform (black/solid curve), the PN waveform (blue/dashed curve), and also the hybrid waveform (orange/dotted curve), while in the bottom panel we plot the relative error between the hybrid waveform and both the NR waveform (black curve) and the PN waveform (blue curve). In each panel, we also show the hybridization interval in purple. The NR waveform in each of these plots is a CCE waveform.\\
	BBH merger: \texttt{q4\_aligned\_chi0\_4} (see Table~\ref{tab:runs}).}
\end{figure}

\begin{figure}
	\label{fig:hybriderror}
	\centering
	\includegraphics[width=\columnwidth]{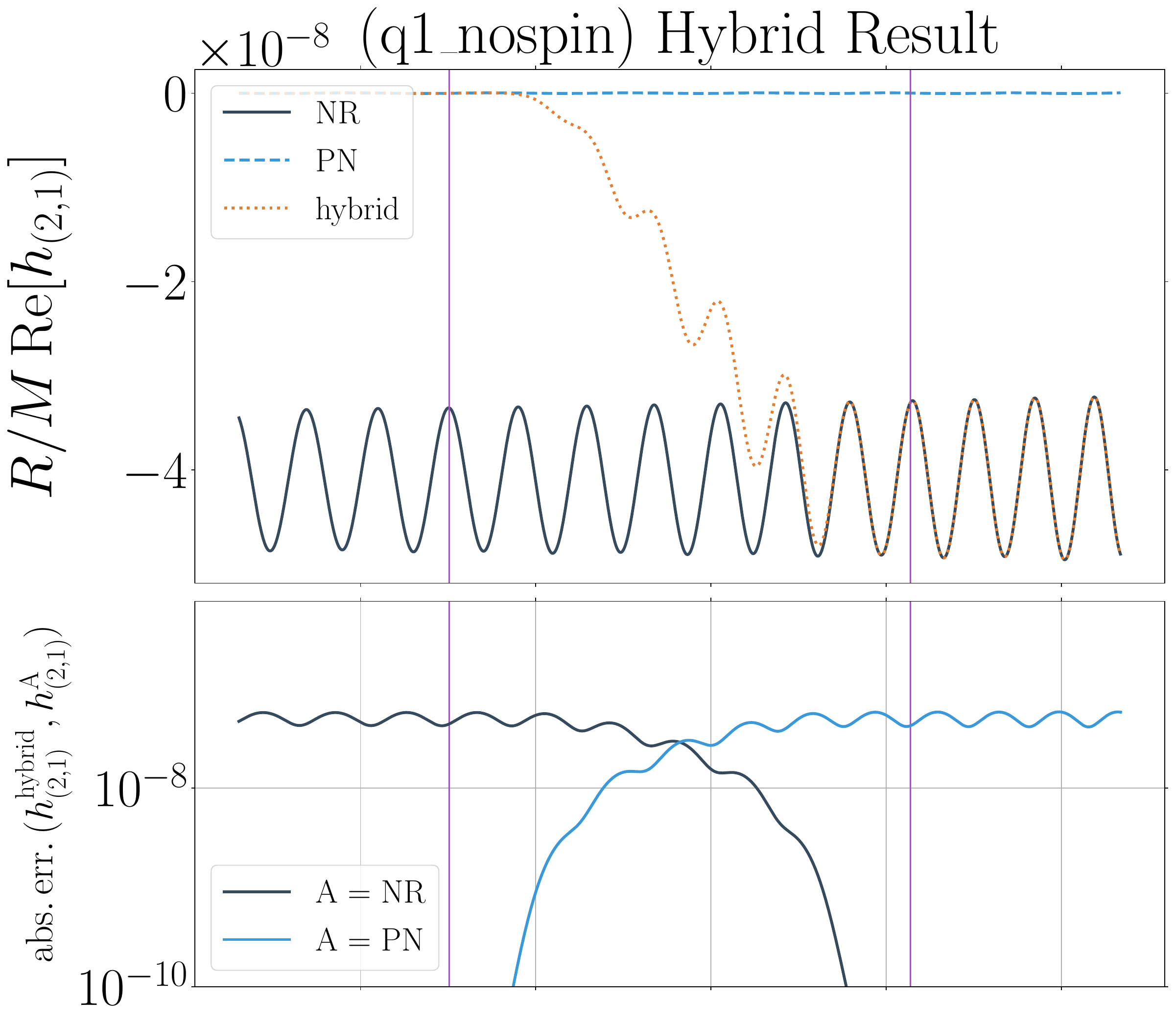}
	\caption{Illustrating a caveat about NR and PN hybridizations, as shown through the strain $(2,1)$ mode. In the top panel we plot the NR waveform (black/solid curve), the PN waveform (blue/dashed curve), and also the hybrid waveform (orange/dotted curve), while in the bottom panel we plot the absolute error between the hybrid waveform and both the NR waveform (black curve) and the PN waveform (blue curve). In each panel, we also show the hybridization interval in purple. See the text for more details. The NR waveform in this plot is a CCE waveform.\\
	BBH merger: \texttt{q1\_nospin} (see Table~\ref{tab:runs}).}
\end{figure}

With our NR waveforms now optimally mapped to the PN BMS frame, we perform strain hybridizations between NR and PN to illustrate the operations this alignment procedure allows for. To create these hybridizations, we use the smoothing function
\begin{align}
f(x)=\begin{cases}
0&x\leq0,\\
\left(1+\exp\left[\frac{1}{x-1}+\frac{1}{x}\right]\right)^{-1}&0<x<1,\\
1&x\geq1.
\end{cases}
\end{align}
so that before the hybridization interval, which is the same as the alignment interval, the hybrid is equal to the PN waveform and after it is equal to the NR waveform. Put differently, we build the hybrid waveform $h^{\text{hybrid}}$ via
\begin{align}
h^{\text{hybrid}}=h^{\text{PN}}+f\left(\frac{u-u_{1}}{u_{2}-u_{1}}\right)\left(h^{\text{NR}}-h^{\text{PN}}\right),
\end{align}
where $u_{2}$ is the time that is four orbits past $u_{1}$.

In Fig.~\ref{fig:hybrid}, we show three plots, one for each of the strain $(2,2)$, $(2,1)$, and $(2,0)$ modes. In each plot, we compare the hybrid waveform to the NR and PN strains for simulation \texttt{q4\_aligned\_chi0\_4}. In each top panel, we show the strain modes, while in each bottom panel, we show the relative error between the hybrid and the NR and PN strain waveforms. As is expected from our alignment results in Fig.~\ref{fig:PNAlignmentErrors}, there is fairly impressive agreement in every mode. Furthermore, the plot of the strain $(2,0)$ mode shows that the initial value of the hybrid now agrees with PN, i.e., it exhibits the memory that we expect to be there due to the emission of radiation throughout the entire past history of the binary's inspiral. Apart from these important positive results, though, there is one minor caveat regarding this hybridization procedure that is worth mentioning to avoid confusion.

\begin{figure}
	\label{fig:comerror}
	\centering
	\includegraphics[width=\columnwidth]{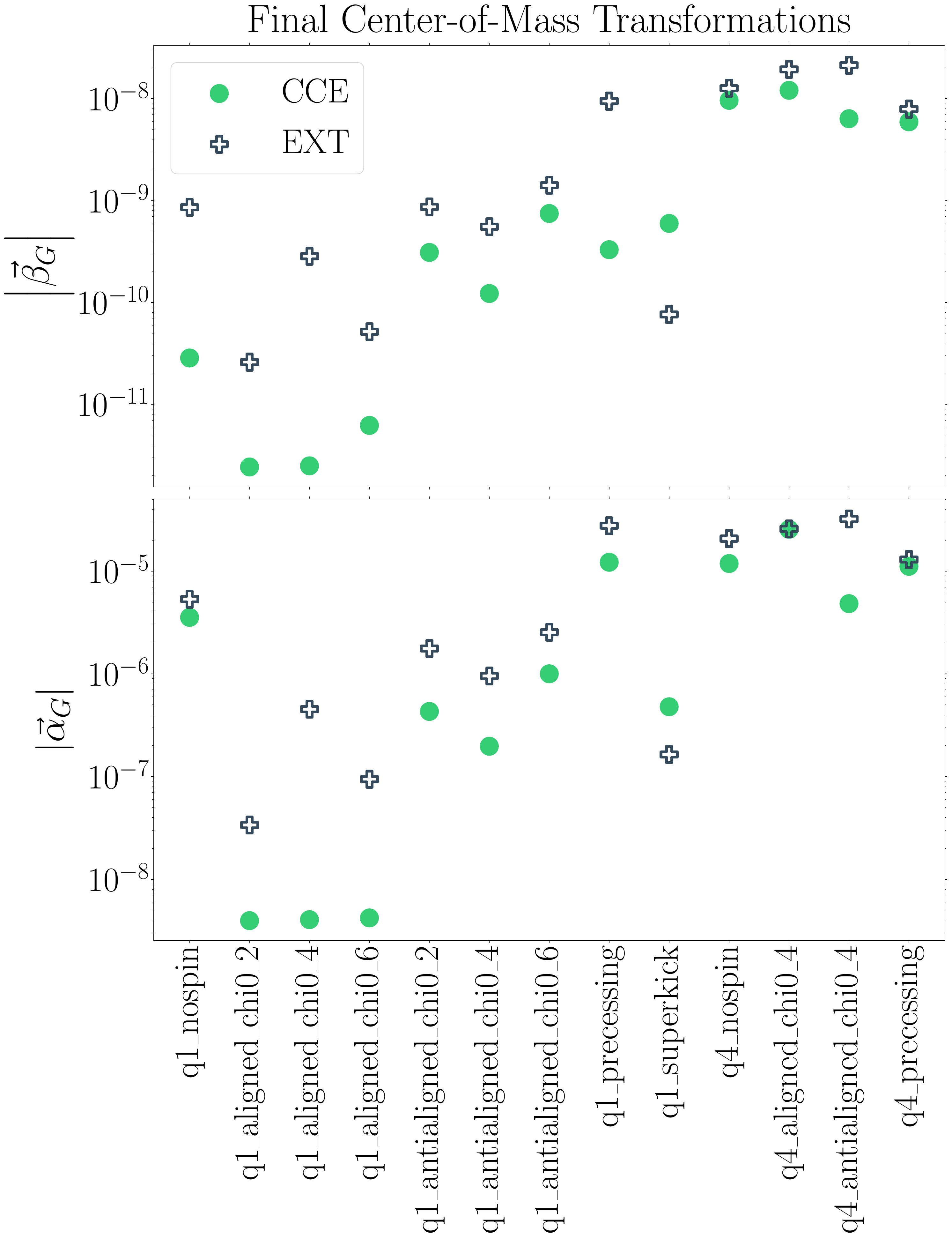}
	\caption{The magnitudes of the boosts and translations needed to map the NR strains in the PN BMS frame to their center-of-mass frame. The reason these are not exactly zero is because a few of the BMS transformations that are not involved in the center-of-mass correction do not commute with the center-of-mass transformation, causing the system to be pushed slightly away from the center-of-mass frame to obtain better alignment with the PN strain waveform. Nonetheless, these transformations are fairly negligible, especially when compared to those in the top plots of Fig.~\ref{fig:boostsandtranslations}.}
\end{figure}

\begin{figure*}
	\label{fig:LevAlignmentErrors}
	\centering
	\includegraphics[width=\textwidth]{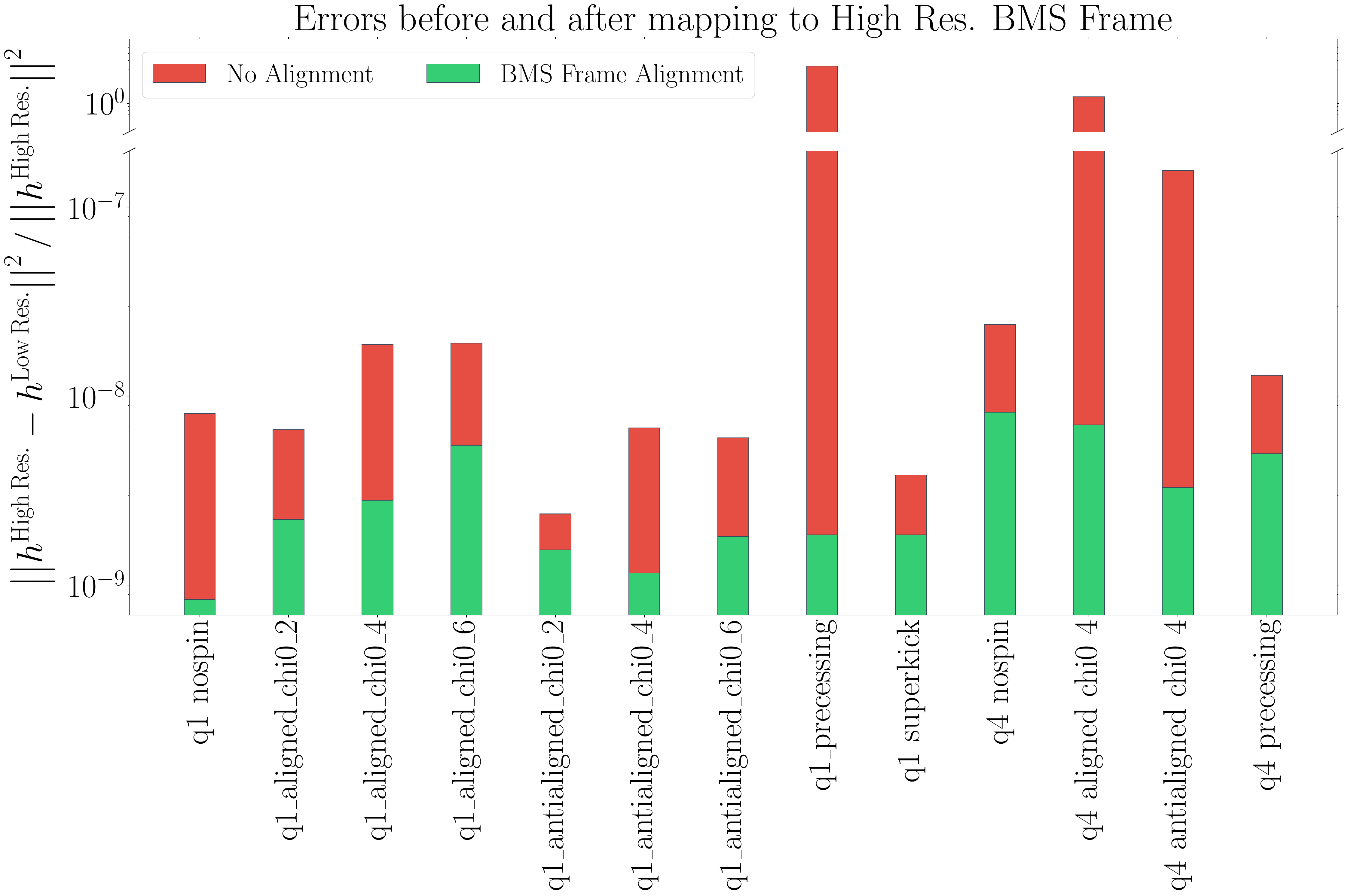}
  \caption{Comparing the BMS alignment results between a high resolution and a lower resolution numerical relativity waveform. In red, we show the initial misalignment, while in green, we show the misalignment after using the new BMS frame alignment: that is, finding the BMS transformation (up to $\ell=4$) that minimizes the average $L^{2}$ norm of the difference of the two waveforms. Note that the measure of this misalignment, $||f(u,\theta,\phi)||^{2}$, is defined to be $||f(u,\theta,\phi)||^{2}=\int_{u_{1}}^{u_{1}+\text{four orbits}}\int_{S^{2}}f\bar{f}\,d\Omega du$.}
\end{figure*}

Recall that when mapping a NR strain waveform to its corresponding PN BMS frame, we apply certain $2\leq\ell\leq4$ supertranslations. Because supertranslations also affect the Bondi time $u$, however, when we supertranslate the strain we not only have to act on it with the supertranslation, but we also have to interpolate the waveform on to a new series of Bondi times. So, when applying a supertranslation with only one nonzero mode, not only will that mode of the strain change, but so will every other mode because of the time interpolation. The reason why this is important is illustrated in Fig.~\ref{fig:hybriderror}.

In Fig.~\ref{fig:hybriderror}, we show the $(2,1)$ mode of the strain hybrid for simulation \texttt{q1\_nospin}. Based purely on symmetry, we would normally expect this mode to be zero, but because of numerical error, the spins and eccentricity of this system are not exactly zero. We therefore observe nonzero values in this mode of the NR waveform, even though they are negligible. Thus, the $(2,1)$ mode of the hybrid looks a bit strange because of this difference between NR and PN.

Apart from this, however, one may also notice that the zero average value of the PN strain is not matched by the nonzero average of the NR strain. This oddity results from the supertranslation's broader influence on the whole of the strain waveform because of the needed time interpolation, as mentioned earlier. In other words, supertranslating the $(2,1)$ mode away from an average of zero improves the error in other modes more than it worsens the error in the $(2,1)$ mode. Consequently, while this kind of behavior in the hybridized waveform is certainly undesirable, it is a natural consequence of the PN waveform not being a perfect theoretical model for the numerical BBH system. Furthermore, we note that this behavior only occurs in modes whose amplitude is rather negligible, i.e., $10^{-6}$ or less.

It is also fairly important to note that, because the center-of-mass transformation does not commute with the other BMS transformations, this procedure of mapping to the PN BMS frame does not exactly map our systems to the center-of-mass frame, even though they are very close. We illustrate this in Fig.~\ref{fig:comerror}, which shows the remaining boosts and translations needed to map the waveform in the PN BMS frame to its center-of-mass frame. As is shown, the necessary boosts and translations are nearly zero, especially when compared to the transformations in the top plots of Fig.~\ref{fig:boostsandtranslations}. Therefore, we consider this mapping to the PN BMS frame to be remarkably successful, especially since the improvements in Fig.~\ref{fig:PNAlignmentErrors} are so notable.

Finally, we show our last result regarding the benefit of utilizing the whole BMS freedom of numerical waveforms. When examining physical quantities that are output by numerical relativity, it is important to run convergence tests to ensure that conclusions can be made with respect to numerical error. In Fig.~\ref{fig:LevAlignmentErrors} we show how convergence tests can be improved by mapping waveforms that are from simulations of different numerical resolutions to the same BMS frame before they are compared with one another. To do this, we perform the same iterative process as described earlier, but we now optimize over every BMS transformation, rather than everything but the center-of-mass transformation. In red, we show the initial misalignment, while in green we show the misalignment after the BMS frame alignment procedure. As is shown, the improvements are relatively minor, but could still prove to be important for numerical simulations run with newer codes, such as SpECTRE~\cite{CodeSpECTRE}, which will be more accurate than the SXS Collaboration's current code SpEC.

\section{Conclusion}

Like any physical system, understanding the frame that a binary black hole merger is in relative to a family of observers is essential. Understanding this frame will help us ensure that any analyses on waveform models are performed properly and no misleading assertions are made. For gravitational-wave physics, fixing the frame is not as simple as fixing the Poincar\'e frame, since the symmetries of asymptotically flat spacetimes are characterized by an infinite extension of the Poincar\'e group: the BMS group.

Currently, gravitational-wave physicists who analyze models of gravitational waves expect those models to be in the center-of-mass frame and the PN Bondi frame, since this is the BMS frame that analytic models are in. However, the waveforms that are currently produced by numerical relativity---the supplier of the most accurate models of gravitational waves---typically are not in such a frame because of an unexpected center-of-mass drift in numerical simulations and a lack of initial data that contains information about the entire past history of the binary black hole's inspiral. Consequently, they are instead in some other BMS frame. But with a proper understanding of the BMS group, one can postprocess these waveforms and map them to the desired BMS frame after the BBH simulation is complete. As of now, such a postprocessing technique is used for the Poincar\'e frame by using the Newtonian center of mass~\cite{Woodford:2019tlo}, but there is no such postprocessing for the Bondi frame.

In this paper, by utilizing asymptotic Poincar\'e charges, i.e., the center-of-mass charge (see Eq.~\eqref{eq:centerofmasscharge}), we show that this method that relies on Newtonian trajectories for mapping to the center of mass Frame is not as successful as previously thought. As a result, we develop an improved procedure for fixing the Poincar\'e frame, which shows large benefits in terms of exhibiting the expected behavior in the asymptotic Poincar\'e charges and also in minimizing certain modes of the strain waveform that are expected to be zero because of axisymmetry.

We also found that we can meaningfully fix the whole BMS frame of our numerical waveforms by mapping them to their corresponding PN BMS frame using a 3PN-order strain waveform. With this BMS frame fixing procedure, we observe that we can produce much more favorable hybridizations between NR and PN strain waveforms than if one were to use an alignment scheme that only utilized the Poincar\'e transformations. Last, we also find that such a BMS frame alignment will prove important for future numerical relativity codes that will be able to run simulations at higher resolutions and will need to properly test the convergence of their waveforms.

With this new method of fully fixing the BMS frame of asymptotic waveforms, many important improvements to gravitational wave modeling can be made. For example, by correctly mapping to the center-of-mass Poincar\'e frame and PN BMS frame, we can produce much better PN and NR hybridizations. Furthermore, because we can now ensure that waveforms are in the same BMS frame, surrogate models built from such waveforms should be more accurate since they are no longer trying to link waveforms in radically different BMS frames. It would be very interesting to see how parameter estimation using a NR surrogate changes depending on whether the waveforms used to build the surrogate have been mapped to a consistent BMS frame, such as the PN BMS frame.

\acknowledgments

Computations were performed with the High Performance
Computing Center and the Wheeler cluster at Caltech. This work was supported in
part by the Sherman Fairchild Foundation and by NSF Grants No.~PHY-2011961,
No.~PHY-2011968, and No.~OAC-1931266 at Caltech, NSF Grants No.~PHY-1912081
and No.~OAC-1931280 at Cornell, and NSF Grant No.~PHY-1806356, Grant No.~UN2017-92945 from the Urania Stott Fund of the Pittsburgh Foundation, and the Eberly research funds of Penn State at Penn State.

\appendix

\section{Fixing the supertranslation freedom with the Moreschi supermomentum}
\label{sec:superrestframe}

As mentioned earlier, an alternative method for fixing the Bondi frame of NR waveforms, apart from mapping to the PN BMS frame, is to map to the super rest frame. Therefore, since this technique could prove rather useful in future analyses---such as quasinormal mode fitting--- we now briefly illustrate how to perform this procedure. According to Moreschi~\cite{Moreschi:1988pc,Moreschi:1998mw}, a reasonable choice for the supermomentum is
\begin{align}
\Psi^{\text{M}}(u,\theta,\phi)\equiv\sum\limits_{\ell\geq0}\sum\limits_{m\leq|\ell|}\Psi_{\ell m}^{\text{M}}(u)Y_{\ell m}(\theta,\phi),
\end{align}
where
\begin{align}
\Psi_{\ell m}^{\text{M}}(u)\equiv-\frac{1}{\sqrt{4\pi}}\int_{S^{2}}Y_{\ell m}\Psi^{\text{M}}(u)\,d\Omega,
\end{align}
and
\begin{align}
\Psi^{\text{M}}(u)\equiv\Psi_{2}+\sigma\dot{\bar{\sigma}}+\eth^{2}\bar{\sigma}.
\end{align}
While this is the supermomentum that correctly defines the Bondi frame, it is important to note that there are many other supermomenta that have been proposed in the literature, e.g., the Bondi-Sachs (BS), Geroch (G), and Geroch-Winicour (GW) supermomenta~\cite{Geroch1977,Geroch:1981ut,Dray:1984rfa,Dain:2002mj}
\begin{subequations}
	\label{eq:supermomenta}
	\begin{align}
	\label{eq:BondiSachs}
	\Psi^{\text{BS}}(u)&\equiv\Psi_{2}+\sigma\dot{\bar{\sigma}},\\
	\label{eq:Geroch}
	\Psi^{\text{G}}(u)&\equiv\Psi_{2}+\sigma\dot{\bar{\sigma}}+\left(\eth^{2}\bar{\sigma}-\bar{\eth}^{2}\sigma\right),\\
	\label{eq:GerochWinicour}
	\Psi^{\text{GW}}(u)&\equiv\Psi_{2}+\sigma\dot{\bar{\sigma}}-\bar{\eth}^{2}\sigma.
	\end{align}
\end{subequations}
While all of these agree on their $\ell<2$ modes,\footnote{The $\ell<2$ modes of the supermomenta in Eq.~\eqref{eq:supermomenta} all correspond to the usual Bondi four-momentum.} the Moreschi supermomentum is the supermomentum that can most easily be used to map to the super rest frame for reasons discussed in~\cite{Moreschi:1988pc,Moreschi:1998mw,Dain:2002mj,Gallo:2014jda}.

It is important to note, however, that
\begin{align}
\label{eq:changeinsupermomentum}
\Delta\Psi^{\text{M}}&\equiv\int_{u_{1}}^{u_{2}}\dot{\Psi}^{\text{M}}(u)\,du\nonumber\\
&=\int_{u_{1}}^{u_{2}}\left[\left(\dot{\Psi}_{2}+\left[\sigma\ddot{\bar{\sigma}}+\eth^{2}\dot{\bar{\sigma}}\right]\right)+|\dot{\sigma}|^{2}\right]du\nonumber\\
&=\int_{u_{1}}^{u_{2}}|\dot{\sigma}|^{2}\,du
\end{align}
since
\begin{align}
\dot{\Psi}_{2}=-\left[\sigma\ddot{\bar{\sigma}}+\eth^{2}\bar{\sigma}\right]
\end{align}
by the Bianchi identities. Consequently, since Eq.~\eqref{eq:changeinsupermomentum} is proportional to the radiated energy, this means that the Moreschi supermomentum can never be made zero so long as there is energy radiated in gravitational waves.\footnote{This also corresponds to the electric part of the null memory~\cite{Mitman:2020pbt}.}
	
With $\Psi^{\text{M}}$, a similar procedure as the one presented in Sec.~\ref{sec:ellgeq2results} for mapping to the PN BMS frame can then be performed. But, instead of minimizing the $L^{2}$ norm of the difference of NR and PN strain waveforms, one would simply find the proper supertranslation that minimizes the $L^{2}$ norm of $\Psi^{\text{M}}$ at a certain time or over some finite time interval, such as the ringdown phase.

\newpage

\bibliography{bibliography}
\end{document}